\documentclass[journal=jacsat,manuscript=article]{achemso}
\pdfoutput=1

\usepackage[version=3]{mhchem} 
\usepackage{algorithm}
\usepackage{algpseudocode, float}
\usepackage{enumitem}
\usepackage{tabls}
\usepackage[table,xcdraw]{xcolor}
\usepackage{booktabs,caption}
\usepackage[flushleft]{threeparttable}
\usepackage{booktabs}
\usepackage{float}
\usepackage{paralist}
\usepackage{multirow}
\SectionNumbersOn

\author{Jiawei Zhan}
\affiliation{Pritzker School of Molecular Engineering,
University of Chicago, Chicago, Illinois 60637, United States}
\author{Marco Govoni}
\affiliation{Pritzker School of Molecular Engineering,
University of Chicago, Chicago, Illinois 60637, United States}
\alsoaffiliation{Materials Science Division and Center for Molecular Engineering, Argonne National Laboratory, Lemont, Illinois 60439, United States}
\alsoaffiliation{Department of Physics, Computer Science, and Mathematics, University of Modena and Reggio Emilia, Modena, 41125, Italy}
\author{Giulia Galli}
\affiliation{Pritzker School of Molecular Engineering,
University of Chicago, Chicago, Illinois 60637, United States}
\alsoaffiliation{Materials Science Division and Center for Molecular Engineering, Argonne National Laboratory, Lemont, Illinois 60439, United States}
\alsoaffiliation{Department of Chemistry, University of Chicago, Chicago, Illinois 60637, United States}
\email{gagalli@uchicago.edu}

\title[An \textsf{achemso} demo]
  {Nonempirical Range-Separated Hybrid Functional with Spatially Dependent Screened Exchange}
\date{\today}

\abbreviations{IR,NMR,UV}

\begin{document}

\begin{abstract}

Electronic structure calculations based on Density Functional Theory  have successfully predicted numerous ground state properties of a variety of molecules and materials.  However,  exchange and correlation functionals currently used in the literature, including semi-local and hybrid functionals, are often inaccurate to describe the electronic properties of heterogeneous solids, especially systems composed of building blocks with large dielectric mismatch. Here, we present a dielectric-dependent range-separated hybrid functional, SE-RSH, for the investigation of heterogeneous materials. We define a spatially dependent fraction of exact exchange inspired by the static Coulomb-hole and screened-exchange (COHSEX) approximation used in many body perturbation theory, and we show that the proposed functional accurately predicts the electronic structure of several non-metallic interfaces, three- and two-dimensional, pristine and defective solids and nanoparticles. 

\end{abstract}

\maketitle

\section{\label{sec1:level1} Introduction}

Materials are heterogeneous systems, often composed of different building blocks including interfaces\cite{mannhar2012interface} and defects. Density Functional Theory (DFT)\cite{10.1103/physrev.136.b864, 10.1103/physrev.140.a1133, teale2022dft} has been extremely successful in predicting ground state and thermodynamic properties of many materials. However, a density functional that provides an equally accurate description of the electronic structure  of systems composed of portions with different dielectric properties is not yet available (for example, a low-band-gap semiconductor interfaced with an insulating oxide, or an amorphous solid interfaced with a crystalline one). In general, density functionals  appropriate to study thermodynamic properties may not accurately predict other quantities, such as electronic excitations and transport coefficients\cite{cohen2012challenges, verma2020status}. In particular, the predictions of simple properties of heterogeneous materials such as band gaps and offsets are still challenging when using DFT\cite{hinuma2014band, ghosh2022efficient}. Hence the development of density functionals for heterogeneous systems remains a critical need of the materials science and condensed matter physics communities.
We emphasize that accurate calculations of electronic band gaps and energy level  alignments in heterostructures are essential to understand the functionality and performance of semiconductor devices\cite{kroemer2001nobel}, as well as of solar and photo-electrochemical cells\cite{gratzel2001photoelectrochemical}.  Band gaps and band offsets are necessary inputs for the evaluation of the quantum efficiency of several processes, including electron-hole separation or charge recombination in semiconductors and insulators.

Importantly, even when using theories beyond mean-field (e.g. many-body perturbation theory\cite{10.1103/physrev.139.a796, martin2016}, dynamical mean field theory\cite{georges1996dynamical, kotliar2006electronic} or Quantum Monte Carlo\cite{becca2017quantum}), DFT results are almost always required as input. Another key reason to improve the accuracy and broaden the applicability of DFT calculations is the need to obtain reliable data to apply machine learning (ML) algorithms, as the success and promise of ML rely on the accuracy of computational and experimental data constituting the training sets\cite{kulik2022roadmap}. 
  
DFT was originally applied to compute the structural and electronic properties of condensed matter systems using the local density approximation\cite{kohn1965self, ceperley1980ground, perdew1981self, perdew1992accurate} (LDA)  and then gradient corrected approximations\cite{10.1103/physrevlett.77.3865,10.1103/physrevb.46.6671} (GGA) to the exchange and correlation energy functional (xc). Subsequently, hybrid functionals\cite{10.1002/wcms.1378, 10.1063/1.464304, 10.1063/1.1528936}, utilized first for finite, molecular systems, were adopted for solids as well. In hybrid DFT, the exchange energy is defined as a linear combination of exact (Hartree-Fock) and local exchange\cite{10.1021/acs.chemrev.8b00193} energies. Among hybrid functionals, PBE0\cite{adamo1999toward} and HSE\cite{10.1063/1.1564060, 10.1063/1.2204597, 10.1063/1.2404663} have been popular ones to describe condensed systems. Recently, dielectric-dependent hybrid\cite{PhysRevB.89.195112, 10.1103/physrevb.83.035119, 10.1021/acs.jctc.7b00368, 10.1103/physrevb.93.235106} functionals have been increasingly adopted to investigate the structural and electronic properties of solids\cite{10.1103/physrevb.83.035119, 10.1021/acs.jctc.7b00368}, liquids\cite{10.1038/s41467-017-02673-z, 10.1126/sciadv.1603210} and also of several molecules\cite{10.1021/acs.jctc.7b00368}. Other emerging approaches include Koopmans-compliant functionals\cite{borghi2014koopmans, 10.1021/acs.jctc.8b00976, nguyen2018koopmans, colonna2022koopmans} and the Localized Orbital Scaling Correction\cite{li2018localized, mahler2022localized}, applicable to both molecules and solids.

However, most of these functionals are often not sufficiently accurate to predict the electronic properties of heterogeneous systems, including surfaces and interfaces, if different portions of the system exhibit substantially different dielectric screening. Several approaches\cite {10.1063/1.4908061,10.1021/acs.jctc.7b00853,10.1103/physrevmaterials.3.073803} have been recently proposed in an attempt to address this problem.  Shimazaki et al.\cite{10.1063/1.4908061} introduced a descriptor of the local environment surrounding atoms in a semiconductor, which they used to define a position-dependent atomic dielectric constant. Borlido et al.\cite{10.1021/acs.jctc.7b00853} defined a local dielectric function using an empirical, system-dependent fitting process. Zheng et al.\cite{10.1103/physrevmaterials.3.073803} introduced instead a nonempirical Screened-Exchange Dielectric-Dependent hybrid (SE-DDH) functional, using intuitive assumptions on local dielectric screening and showing promising results for both heterogeneous semiconductors and insulators.

In this work, we introduce a non-empirical, range-separated, dielectric-dependent functional which provides consistent and accurate predictions of the electronic properties of both 3D and 2D solids, as well as of interfaces and finite systems. The functional is based on the definition of a mixing fraction of exact and local exchange energies and potentials, inspired by the static COulomb Hole and Screened-EXchange (COHSEX) approximation\cite{PhysRevLett.55.1418} used in many body perturbation theory. We present a thorough validation of our results for various condensed systems, with focus on fundamental electronic properties such as band gaps and energy level alignments.\\
\indent The remainder of this paper is organized as follows: In Sec. \ref{sec1:level2}, we derive the expression of the Screened-Exchange Range-Separated Hybrid (SE-RSH) functional and establish a connection with more advanced electronic structure methods. We then present applications to the electronic structure of a variety of complex materials in Sec. \ref{sec3}. We conclude with Sec. \ref{sec:4}.

\section{\label{sec1:level2}Method}
\subsection{General hybrid functionals for complex materials}

The generalized Kohn-Sham (GKS) nonlocal potential $v_{\rm{GKS}}(\mathbf{r,r'})$ entering the Kohn-Sham (KS) Hamiltonian is given by:
\begin{equation}
    \begin{aligned}
        v_{\rm{GKS}}(\mathbf{r, r'}) = v_{H}(\mathbf{r}) + v_x(\mathbf{r, r'}) + v_c(\mathbf{r}) + v_{\rm{ext}}(\mathbf{r}), \label{equ.1}
    \end{aligned}
\end{equation}
where $v_{H}(\mathbf{r})$, $v_{x}$ and $v_c(\mathbf{r})$ are the Hartree, nonlocal exchange and correlation potential, respectively, and $v_{\mathrm{ext}}(\mathbf{r})$ is the attractive Coulomb potential between electrons and nuclei. In hybrid DFT, the term $v_{x}(\mathbf{r,r'})$ is given by a linear combination of exact and semilocal exchange potential, and the mixing fraction $\alpha(\mathbf{r, r'})$ between the two depends on the specific hybrid functional. In Range-Separated Hybrid (RSH) functionals\cite{leininger1997combining, doi:10.1063/1.2409292, vydrov2006importance, yanai2004new, chai2008long, stein2009reliable, baer2010tuned}, $\alpha(\mathbf{r, r'})$ has the following form:
\begin{equation}
    \alpha(\mathbf{r, r'}) = m + (n - m)\rm{erfc}\left(\mu|\mathbf{r-r'}|\right), \label{equ:2}
\end{equation}
where the values of $m$ and $n$ determine the fraction of exact exchange included in the long-range (lr) and short-range (sr) components, respectively. The screening parameter $\mu$ defines how lr and sr components are connected to each other. By using $\alpha$ defined above, $v_{x}(\mathbf{r,r'})$ is:
\begin{equation}
\begin{aligned}
    v_{x}(\mathbf{r,r'}) = &\alpha(\mathbf{r, r'})\Sigma_{x}(\mathbf{r, r'}) \\
    & + (1 - m)v_{x}^{\rm{lr}}(\mathbf{r};\mu) + (1 - n)v_{x}^{\rm{sr}}(\mathbf{r};\mu), \label{equ:3}
\end{aligned}
\end{equation}
where the Hartree-Fock exchange $\Sigma_x(\mathbf{r, r'})$ is given by:
\begin{equation}
    \Sigma_x(\mathbf{r, r'}) = -\sum_{j}^{N}\phi_{j}^*(\mathbf{r'})\phi_{j}(\mathbf{r})v(\mathbf{r,r'}), \label{equ:4}
\end{equation}
and $\phi_{j}$ is the $j^{\text{th}}$ KS single-particle wave function, $N$ is the number of occupied electronic states and $v(\mathbf{r, r'})$ is the Coulomb potential. The lr and sr components of the semilocal exchange potential, denoted as $v_{x}^{\rm{lr}}(\mathbf{r};\mu)$ and $v_{x}^{\rm{sr}}(\mathbf{r};\mu)$ respectively, depend only on the density $\rho(\mathbf{r})$ and its gradient. Here we adopt the PBE\cite{10.1103/physrevlett.77.3865} approximation for the semilocal exchange $v_{x}(\mathbf{r})$ and correlation $v_{c}(\mathbf{r})$.
\\ \indent The expression of many of the exchange functionals commonly used in the literature, including PBE0\cite{adamo1999toward} and HSE \cite{10.1063/1.1564060, 10.1063/1.2204597, 10.1063/1.2404663}, may be recovered from Eq.(\ref{equ:3}). In some recently defined Range-Separated Dielectric-Dependent Hybrid functionals (RS-DDH)\cite{10.1103/physrevb.93.235106, 10.1103/physrevmaterials.2.073803}, the values of ${m, n, \mu}$ are related to  system-dependent dielectric properties computed from first principle. These functionals were shown to yield accurate electronic properties of inorganic materials and molecular crystals. 
\\ \indent However, in the case of RS-DDH functionals, the mixing fraction entering Eq.(\ref{equ:3}) is not expected to be appropriate to describe interfaces and surfaces. Recently, Zheng et al.\cite{10.1103/physrevmaterials.3.073803} introduced a mixing fraction defined using a local, spatially-dependent dielectric function $\epsilon(\mathbf{r})$:
\begin{equation}
    \alpha^{\rm{SE-DDH}}(\mathbf{r, r'}) = \frac{1}{\sqrt{\epsilon(\mathbf{r})\epsilon(\mathbf{r'})}}. \label{equ: 5}
\end{equation}
Building on Ref.~\citenum{10.1103/physrevmaterials.3.073803}, we define a Screened-Exchange Range-Separated Hybrid (SE-RSH) functional that extends the applicability of RS-DDH to complex, heterogeneous materials, and we define $\alpha(\mathbf{r, r'})$ as follows:
\begin{eqnarray}
    \alpha^{\rm{SE-RSH}}(\mathbf{r, r'}) = && \left(1-\frac{1}{\sqrt{\epsilon(\mathbf{r})\epsilon(\mathbf{r'})}}\right)\rm{erfc}\left(\mu(\mathbf{r})|\mathbf{r-r'}|\right) \nonumber \\
    && + \frac{1}{\sqrt{\epsilon(\mathbf{r})\epsilon(\mathbf{r'})}}.\label{equ: alpha}
\end{eqnarray}
Here, $\epsilon(\mathbf{r})$ and $\mu(\mathbf{r})$ represent a local dielectric function and local screening function, respectively, obtained from first principles as discussed below (Section \ref{sec1:localfunction}). Table \ref{tab:difffunc} summarizes the mixing fraction of exact exchange used to define all energy functionals mentioned above.
\begin{table}[htbp!]
    \caption{The fraction of exact exchange used in several exchange-correlation functionals listed in column 1; $m$, $n$ and $\mu$ denote the long-range, short-range fraction of exact exchange and the screening parameter (function), respectively  (see Eq. \ref{equ:2}). $\epsilon_{\infty}$ is the macroscopic dielectric constant. See Sec. \ref{sec1:localfunction} for the definitions of $\mu(\mathbf{r})$ and $\epsilon(\mathbf{r})$.}
\label{tab:difffunc}
\begin{threeparttable}
\renewcommand{\arraystretch}{1.2} 
\tabcolsep=4pt 
\begin{tabular}{l c c c }
\toprule
  & $m$ & $n$ & $\mu$ \\
\midrule
PBE\cite{10.1103/physrevlett.77.3865} & 0 & 0 & 0 \\
HSE06\cite{10.1063/1.1564060, 10.1063/1.2204597} & 0 & 0.25 & 0.11 \\
DDH\cite{PhysRevB.89.195112} & $1/\epsilon_{\infty}$ & $1/\epsilon_{\infty}$ & 0 \\
RS-DDH\cite{10.1103/physrevb.93.235106} & $1/\epsilon_{\infty}$ & 0.25 & $\mu$ \\
DD0-RSH-CAM\cite{10.1103/physrevmaterials.2.073803} & $1/\epsilon_{\infty}$ & 1 & $\mu$ \\
SE-DDH\cite{10.1103/physrevmaterials.3.073803} & $1/\sqrt{\epsilon(\mathbf{r})\epsilon(\mathbf{r'})}$ & $1/\sqrt{\epsilon(\mathbf{r})\epsilon(\mathbf{r'})}$ & 0 \\
SE-RSH & $1/\sqrt{\epsilon(\mathbf{r})\epsilon(\mathbf{r'})}$ & 1 & $\mu(\mathbf{r})$ \\
\bottomrule
\end{tabular}
\renewcommand{\arraystretch}{1} 
\end{threeparttable}
\end{table}
\\ \indent Using Eq.(\ref{equ: alpha}), the exchange and correlation potential of SE-RSH is then expressed as:
\begin{equation}
\begin{aligned}
    v_{xc}^{\rm{SE-RSH}}\left(\mathbf{r, r'}\right) = &\alpha^{\rm{SE-RSH}}\left(\mathbf{r, r'}\right)\Sigma_{x}\left(\mathbf{r, r'}\right) \\
    & + \left(1-\frac{1}{\epsilon(\mathbf{r})}\right)v_{x}^{\rm{lr}}\left(\mathbf{r};\mu(\mathbf{r})\right) + v_{c}(\mathbf{r}). \label{eq.7}
\end{aligned}
\end{equation}
The lr component of the PBE semilocal exchange can be calculated by scaling the PBE exchange hole\cite{doi:10.1063/1.476928}, $J^{\mathrm{PBE}}\left(s, y\right)$, by the lr screening factor:
\begin{equation}
\begin{aligned}
    v_{x}^{\mathrm{lr}}\left(\mathbf{r};\mu(\mathbf{r})\right) = &-\epsilon_{x}^{\rm{LDA}}\left(\rho(\mathbf{r})\right)\frac{8}{9}\\
    &\times\int_{0}^{\infty}\mathrm{d}y\ yJ^{\mathrm{PBE}}\left(s, y\right)\mathrm{erf}\left(\frac{\mu(\mathbf{r})y}{k_{F}}\right),
\end{aligned}
\end{equation}
where $s=|\nabla \rho|/2k_{F}\rho$ is the reduced gradient, $k_{F}=(3\pi^2\rho)^{1/3}$ and $\epsilon_{x}^{\rm{LDA}}$ is the LDA exchange energy density.\\
\indent As we remark in the Supporting Information, the calculation of the exchange energy entering the SE-RSH functional scales as the number of points used to represent $\mu(\mathbf{r})$. Hence, we adopt a coarse-grained approximation to $\mu(\mathbf{r})$ for computational convenience. Further, an integrable divergence appears as a singularity when the nonlocal exchange energy is formulated in reciprocal space and is present for any strictly positive mixing fraction $\alpha(\mathbf{r, r'})$. This divergence is treated following the procedure proposed by Gygi and Baldereschi\cite{gygi1986self} (see the Supporting Information for details).

\subsection{Local Dielectric and Screening Functions\label{sec1:localfunction}}
 
 We describe next a non-empirical approach to compute the two local functions entering Eq.(\ref{equ: alpha}), namely the local dielectric function $\epsilon(\mathbf{r})$ and the local screening function $\mu(\mathbf{r})$. The former is obtained by DFT calculations in a finite electric field by minimizing the functional\cite{PhysRevB.73.075121, 10.1103/physrevmaterials.3.073803, PhysRevLett.89.117602}:
 \begin{equation}
 \begin{aligned}
     F(\mathbf{E}, [\rho])
     &= E_{\mathrm{KS}}[\rho] - \int \mathbf{E}\cdot\mathbf{r}\rho(\mathbf{r})d\mathbf{r},
 \end{aligned}
 \end{equation}
 where $\int \mathbf{E}\cdot\mathbf{r}\rho(\mathbf{r})d\mathbf{r}$ is called the electric enthalpy, $\mathbf{E}$ is an external electric field and $E_{\mathrm{KS}}$ is the Kohn-Sham energy of the system. In periodic non-metallic systems, the spatially dependent induced polarization due to the field $\mathbf{E}$ can be defined as\cite{10.1103/physrevb.73.075121}:
\begin{equation}
    \Delta P(\mathbf{r}) = -e\sum_{i}^{N}\Delta \mathbf{r}_c^{i}\delta(\mathbf{r}-\mathbf{r}_{c}^{i}),
\end{equation}
where $\Delta \mathbf{r}_c^{i}$ is the shift of
the center ($\mathbf{r}_c$) of the $i^{\text{th}}$ Wannier function induced
by the applied electric field. The spatially dependent $\epsilon$ is then determined as:
\begin{equation}
    \epsilon_{kl}(\mathbf{r}) = \delta_{kl} + 4\pi\frac{\Delta P_{k}(\mathbf{r})}{\Delta E_{l}} \label{epsilon},
\end{equation}
where $k$ and $l$ are Cartesian coordinates.
\begin{figure}[htbp!]
\includegraphics[width=8.6cm]{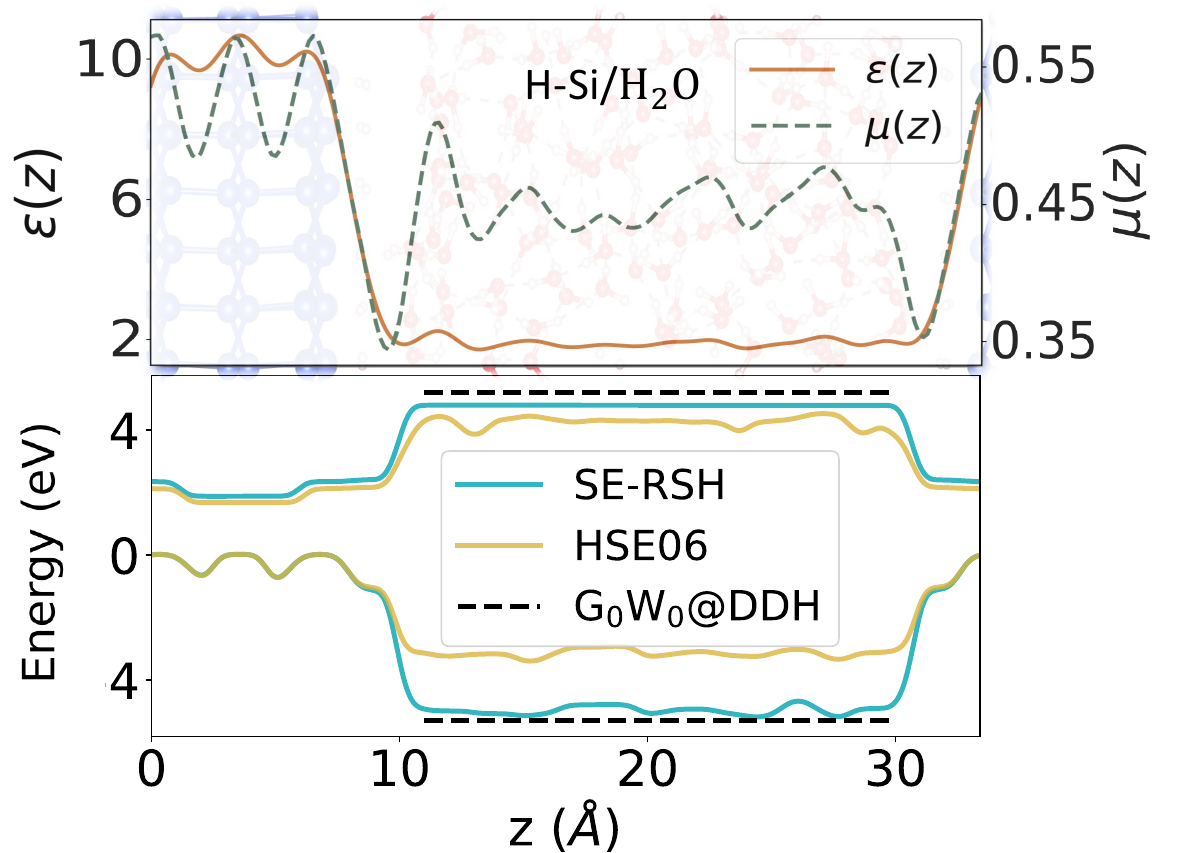}
\caption{\label{fig:1} Top panel: The local dielectric function $\epsilon(z)$ and local screening function $\mu(z)$ for a model H-$\rm{Si/H_2O}$ interface, computed as averages of $\epsilon(\mathbf{r})$ (Eq. \ref{epsilon}) and $\mu(\mathbf{r})$ (Eq. \ref{mu}) in the (x,y) plane parallel to the interface. The z axis is perpendicular to the interface. Bottom panel: Band edges of H-$\rm{Si/H_2O}$ computed (Eq. \ref{ldos}) at different levels of theory, using the SE-RSH functional proposed in this work, the HSE06 functional\cite{10.1063/1.1564060, 10.1063/1.2204597} and $G_0W_0$@DDH 
calculations\cite{gaiduk2018electron}.}
\end{figure}
\\ \indent The local screening function, which generalizes the definition of constant screening parameter $\mu_{\mathrm{TF}}$ introduced by Skone et al.\cite{10.1103/physrevb.93.235106} for homogeneous materials, is related to the volume occupied by the valence electrons:
\begin{equation}
    \mu(\mathbf{r}) = \frac{1}{2}k_{\mathrm{TF}}(\mathbf{r}) = \left(\frac{3\rho_{v}(\mathbf{r})}{\pi}\right)^{1/2}\label{mu},
\end{equation}
where $k_{TF}$ is the Thomas-Fermi screening length and $\rho_{v}$ is the valence electron density. Fig.(\ref{fig:1}) shows $\epsilon(\mathbf{r})$ and $\mu(\mathbf{r})$ computed at the PBE level of theory for a model H-Si/$\mathrm{H_2O}$ interface. We can clearly observe two distinct average values of $\epsilon(\mathbf{r})$ and $\mu(\mathbf{r})$ in the two bulk regions where $\epsilon$ and $\mu$ oscillate around constant values, equal to those one would obtain in the respective bulk systems represented by slabs of the same size.
\\ \indent Alternative definitions of screening parameters for homogeneous materials have been suggested in the literature. For example, Skone et al.\cite{10.1103/physrevb.93.235106} and Chen et al.\cite{10.1103/physrevmaterials.2.073803} defined screening parameters $\mu_\mathrm{fit}$ by using a model to fit the long-range decay of the diagonal elements of the dielectric function [$\epsilon^{-1}$]. However, these authors noted that RS-DDH utilizing either $\mu_{\mathrm{TF}}$ or $\mu_{\mathrm{fit}}$ yields comparable accuracy, with $\mu_{\mathrm{TF}}$ being less computationally demanding to obtain.\\
\begin{figure}[htbp!]
\includegraphics[width=7cm]{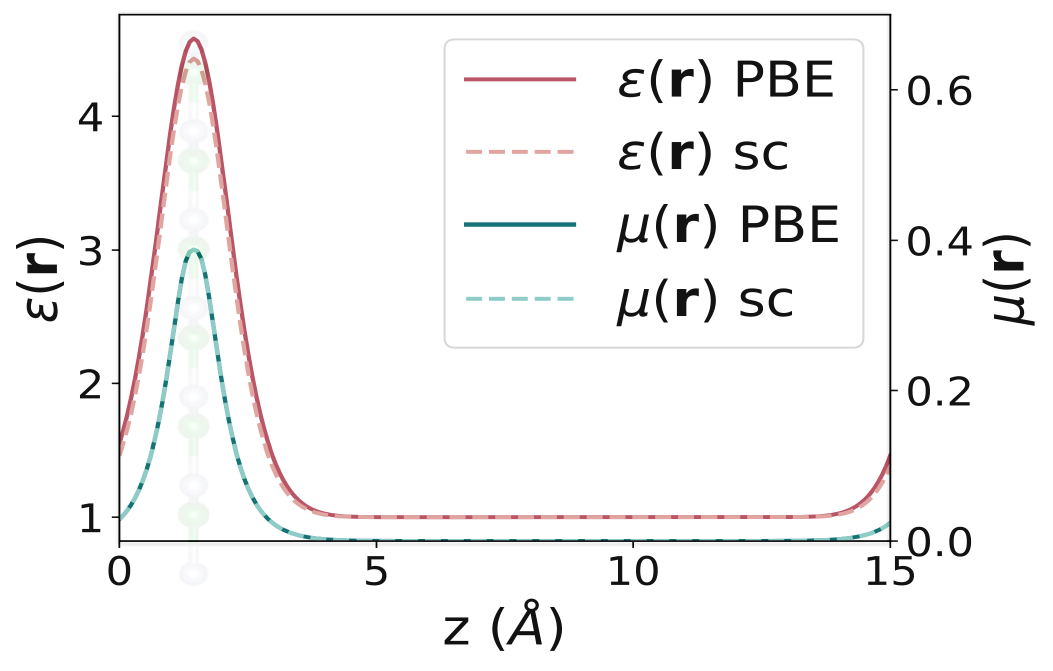}
\caption{\label{fig:sc_local}
 The local dielectric function $\epsilon(z)$ and local
screening function $\mu(z)$ for a h-BN monolayer, computed as averages of $\epsilon(\mathbf{r})$ (Eq. \ref{epsilon}) and $\mu(\mathbf{r})$ (Eq. \ref{mu}) in the (x,y) plane of the monolayer. The z axis is perpendicular to the monolayer. The solid and dashed lines represent calculations conducted at the PBE level of theory and self-consistent calculations (see text), respectively.}
\end{figure}
\indent We have tested the impact of calculating $\epsilon(\mathbf{r})$ and $\mu(\mathbf{r})$ in a self-consistent manner. Specifically, the local functions given by Eq.(\ref{epsilon}) and Eq.(\ref{mu}) are evaluated iteratively until convergence is achieved, akin to the method proposed in Ref.~\citenum{PhysRevB.89.195112, 10.1103/physrevb.93.235106}. A comparison of non-self-consistent (at the PBE level) and self-consistent (sc) $\epsilon(\mathbf{r})$ and $\mu(\mathbf{r})$ for a monolayer h-BN is presented in Fig.(\ref{fig:sc_local}). We also report the fundamental band gap of the system computed with SE-RSH in Fig.(\ref{fig:hbn_sc}), along with $G_0W_0$ results. 

\begin{figure}[htbp!]
\includegraphics[width=8.6cm]{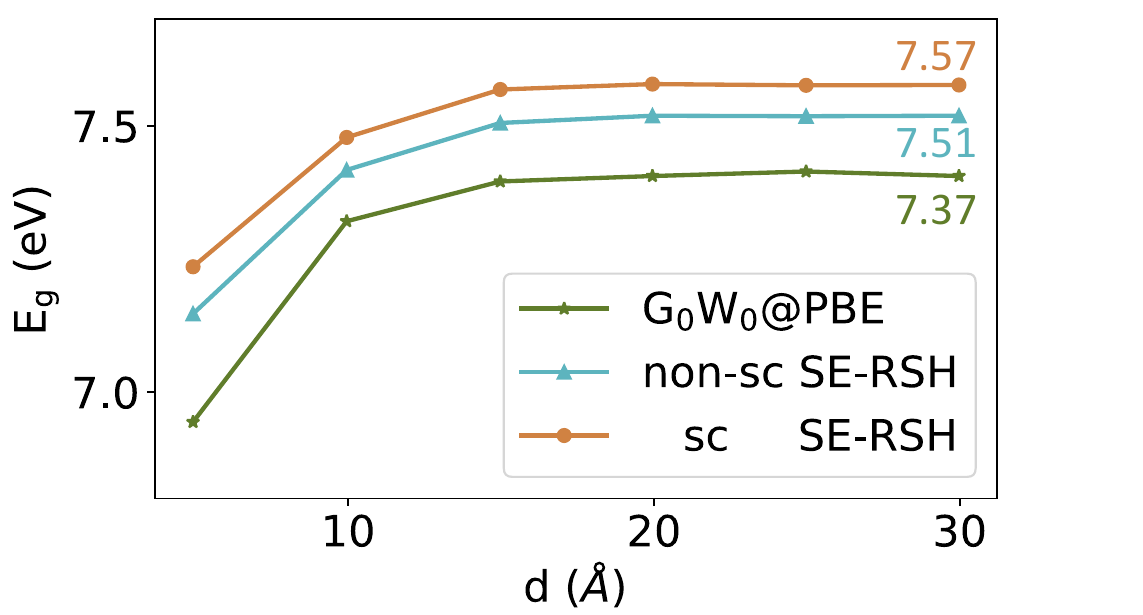}
\caption{\label{fig:hbn_sc} The fundamental band gap ($\mathrm{E_g}$) of a h-BN monolayer, computed as a function of the vacuum length (d) in the supercell. Results are shown using the SE-RSH functional proposed in this work, where the local functions $\epsilon(\mathbf{r})$ and $\mu(\mathbf{r})$ (see Fig. \ref{fig:sc_local}) are computed either at the PBE level of theory (non-sc) or self-consistently (sc). We also show $G_0W_0$ results obtained by truncating the Coulomb potential in two-dimensions\cite{huser2013quasiparticle}.}
\end{figure}

Consistent with previous studies\cite{10.1103/physrevmaterials.3.073803}, we found that utilizing a self-consistent approach for the computation of $\mu(\mathbf{r})$ and $\epsilon(\mathbf{r})$  yields only minor changes in their values, while significantly increasing the computational cost. Therefore, in the following, we report results obtained by the SE-RSH functional with $\epsilon(\mathbf{r})$ and $\mu(\mathbf{r})$ computed at the PBE level of theory, non self-consistently.\\
\indent We note that we have not used the SE-RSH functional for structural relaxations, as an efficient strategy to update $\epsilon(\mathbf{r})$ and $\mu(\mathbf{r})$ when atomic positions vary is still under investigation. Therefore, in our study, all atomic structures were optimized with the PBE functional and then electronic structure calculations were carried out with the SE-RSH functional. However, we note that some studies\cite{10.1103/physrevmaterials.2.073803} have utilized the RS-DDH functional to obtain the lattice constants of homogeneous semiconductors and insulators, including some of those reported in Table \ref{tab:3Dsystems}. For example, Ref.~\citenum{10.1103/physrevmaterials.2.073803} showed that the RS-DDH functional yields lattice constants in better agreement with experiments than the semilocal PBE functional. These results are encouraging since the SE-RSH functional reduces to RS-DDH in homogeneous bulk systems. 

\subsection{Screened Fock Exchange as a simplified Self-Energy}

\begin{figure}[htbp!]
\includegraphics[width=8.6cm]{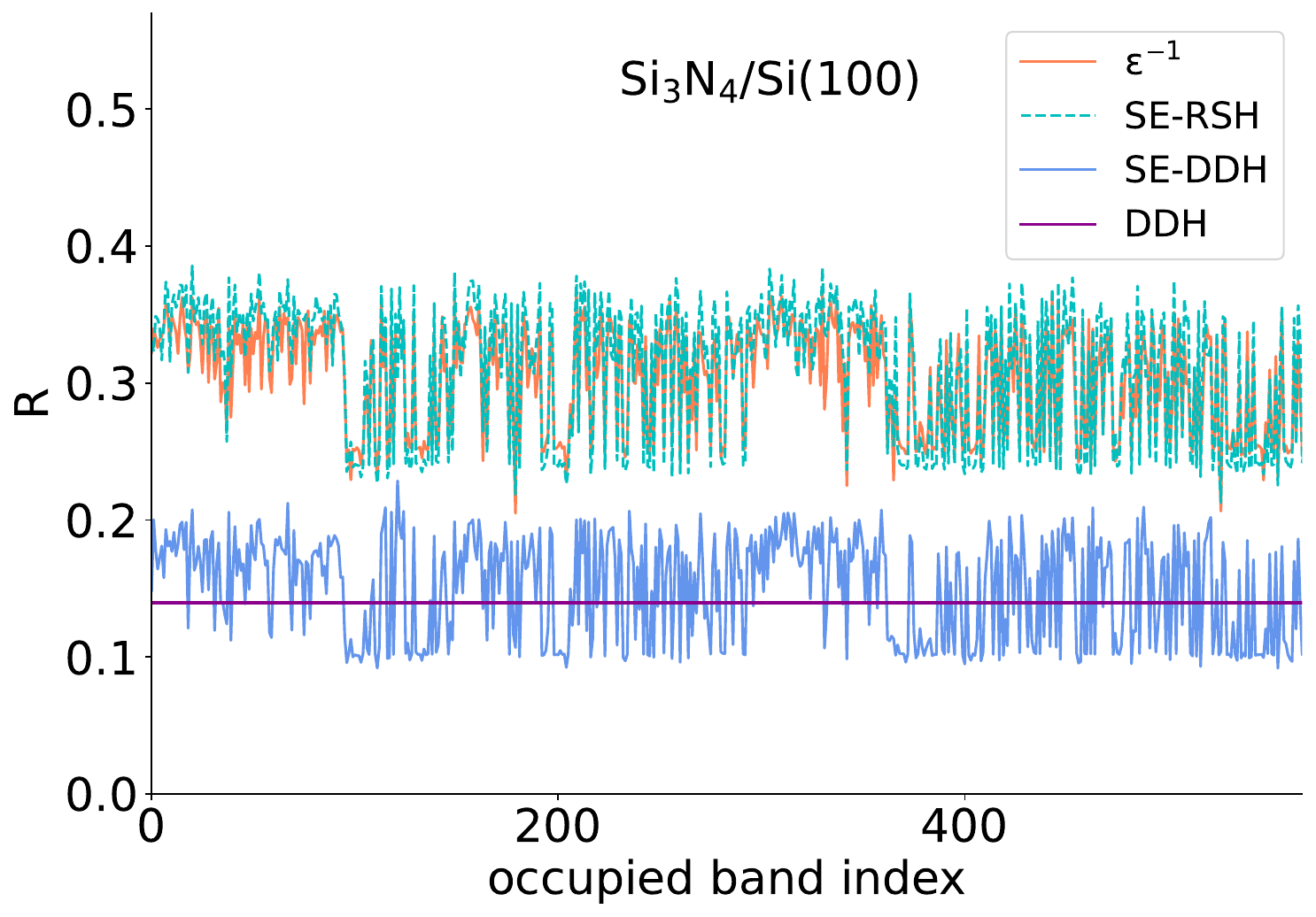}
\caption{\label{fig:2} The state dependent screened-exchange Ratio (R) computed for a model $\mathrm{Si_3N_4}$/Si(100) interfaces, where Si is crystalline and $\mathrm{Si_3N_4}$ is amorphous. We show R computed at different levels of the theory, using Eq.(\ref{SER}) and the functionals SE-DDH\cite{10.1103/physrevmaterials.3.073803}, DDH\cite{PhysRevB.89.195112} and SE-RSH proposed in this work. The results for $\epsilon^{-1}$ were obtained with the \texttt{WEST} code\cite{govoni2015large, yu2022gpu}.}
\end{figure}

We now turn to discuss the connection of the KS equations with the functional defined in the previous section, and the Hedin's equations\cite{10.1103/physrev.139.a796} of many body perturbation theory. In the Hedin's equations,  instead of an exchange-correlation potential, a nonlocal and energy-dependent operator is present, known as the self-energy $\Sigma$. Here we compare the GKS potential of Eq.(\ref{equ.1}) with the self-energy expressed within the static $GW$ approximation\cite{PhysRevLett.55.1418} (static-COHSEX):
\begin{equation}
\Sigma(\mathbf{r, r'}; \omega=0) = \Sigma_{\mathrm{SEX}}(\mathbf{r, r'}) + \Sigma_{\mathrm{COH}}(\mathbf{r, r'}),
\end{equation}
where  $\Sigma_{\mathrm{SEX}}$ is the statically screened-exchange (SEX) and the local $\Sigma_{\mathrm{COH}}$ represents the Coulomb-hole (COH) interaction:
\begin{equation}
    \begin{aligned}
    \Sigma_{\mathrm{SEX}}(\mathbf{r, r'}) &= -\sum_{i=1}^{N}\phi_{i}^*(\mathbf{r'})\phi_{i}(\mathbf{r})W(\mathbf{r, r'})\\
    \Sigma_{\mathrm{COH}}(\mathbf{r, r'}) &= -\frac{1}{2}\delta(\mathbf{r-r'})\left[v(\mathbf{r, r'}) - W(\mathbf{r, r'})\right].
    \end{aligned}\label{equ: 13}
\end{equation}
In Eq.(\ref{equ: 13}), the screened Coulomb potential $W$ is:
\begin{equation}
    W(\mathbf{r, r'}) = \int \mathrm{d}\mathbf{r''}\epsilon^{-1}(\mathbf{r, r''})v(\mathbf{r'', r'}), \label{equ: W}
\end{equation}
where $\epsilon^{-1}$ is the inverse dielectric response function.
\\ \indent The use of the COHSEX's self-energy in the KS equations, instead of the exchange-correlation (xc) potential, has been previously explored by several authors\cite{10.1103/physrevb.93.235106,10.1103/physrevmaterials.2.073803} for homogeneous systems. Below, we compare the results of COHSEX and the SE-RSH xc potential for a representative heterogeneous system.  
\\ \indent We define a screened Hartree-Fock exchange $\Sigma_{x}^{\alpha}$ operator: 
\begin{equation}
    \Sigma_{x}^{\alpha}(\mathbf{r, r'}) = - \sum_{j}^{N}\phi_{j}^*(\mathbf{r'})\phi_{j}(\mathbf{r})\alpha(\mathbf{r, r'})v(\mathbf{r, r'}).
\end{equation}
which depends on the density matrix ($\rho$) and  it is thus invariant under unitary transformations of the occupied manifold. We express $\Sigma_{x}^{\alpha}$ using maximally localized orbitals ($\psi$) obtained from a unitary transformation of the KS eigenstates. Within such a representation , the off-diagoonal elements of the screened Hartree-Fock exchange can be neglected, and the matrix elements of $\Sigma_x^\alpha$ become:
\begin{eqnarray}
    \langle k|\Sigma_{x}^{\alpha}|l \rangle && = -\sum_{j}^{N}\iint \mathrm{d}\mathbf{r}\mathrm{d}\mathbf{r'}\rho_{kj}(\mathbf{r})\alpha(\mathbf{r, r'})v(\mathbf{r, r'})\rho^{*}_{lj}(\mathbf{r'}) \nonumber \\
    && \approx -\langle \rho_{kk}|\alpha \odot v |\rho_{ll}\rangle \delta_{kl},
\end{eqnarray}
where $\rho_{kj}(\mathbf{r})=\psi^{*}_k(\mathbf{r})\psi_j(\mathbf{r})$, $\rho_{kk}(\mathbf{r})$ is the $k^{\mathrm{th}}$ orbital density and $\odot$ is the Hadamard product. In a similar fashion, the matrix elements of $\Sigma_{\mathrm{SEX}}$ in a localized representation become:
\begin{equation}
    \langle k | \Sigma_{\mathrm{SEX}} | l \rangle \approx -\langle \rho_{kk}|\epsilon^{-1}v|\rho_{ll}\rangle\delta_{kl}.
\end{equation}
\indent We then define state-dependent screened-exchange Ratios (R) using the matrix elements of $\Sigma_{x}^{\alpha}$ and $\Sigma_{\mathrm{SEX}}$, respectively:
\begin{equation}
    \begin{aligned}
        \mathrm{R}_{i}^{\alpha} &=\frac{\langle \rho_{ii}|\alpha\odot v|\rho_{ii}\rangle}{\langle \rho_{ii}|v|\rho_{ii}\rangle} \\
        \mathrm{R}_{i}^{\rm{\epsilon^{-1}}} &=\frac{\langle \rho_{ii}|\epsilon^{-1}v|\rho_{ii}\rangle}{\langle \rho_{ii}|v|\rho_{ii}\rangle}.
    \end{aligned} \label{SER}
\end{equation}
\indent We use the Projective Dielectric Eigenpotential Method (PDEP)\cite{10.1103/physrevb.78.113303} to evaluate the static dielectric response function in a separable form\cite{10.1021/ct500958p} and to compute $\mathrm{R}^{\mathrm{\epsilon^{-1}}}$.   
Fig.(\ref{fig:2}) shows $\mathrm{R}^{\rm{\epsilon^{-1}}}$ and $\mathrm{R}^{\rm{\alpha}}$ evaluated for mixing fractions $\alpha(\mathbf{r, r'})$ corresponding to different hybrid functionals for a model $\mathrm{Si_3N_4/Si(100)}$ interface.

In the case of the DDH functional where $\alpha$ = $1/\epsilon_{\infty}$, $\mathrm{R}$ is a constant. The behavior of $\mathrm{R}$ evaluated for the SE-DDH functional is similar to that of  $\mathrm{R^{\rm{\epsilon^{-1}}}}$. However, the absolute values of $\mathrm{R}$ obtained with SE-DDH differ from those of $\mathrm{R^{\rm{\epsilon^{-1}}}}$. Interestingly, when using the range-separated functional SE-RSH, we finally obtain values of $\mathrm{R}$ in excellent agreement with those of $\mathrm{\epsilon^{-1}}$. Hence, we conclude that the mixing fraction used in the definition of SE-RSH is an accurate approximation of the statically screened COHSEX.

\subsection{Computational details}
We implemented the SE-RSH hybrid functional in  the \texttt{Qbox}\cite{gygi2008architecture} code and performed calculations for several systems using supercells and the $\Gamma$ point sampling of the Brillouin zone, with optimized norm-conserving Vanderbilt pseudopotentials\cite{hamann2013optimized}. The energy cutoff for the plane-wave basis set is set at 60 Ry. For two-dimensional systems, we used the lattice parameters presented in Table \ref{tab:mono_lattice} and 6$\times$6 supercells, without performing any additional relaxation of the cell volume. The atomic positions were optimized using the PBE functional until the atomic forces were below 0.05 eV/$\AA$. To minimize the interactions between neighboring supercells, a minimum of 15 $\AA$ vacuum spacing was included in our supercells.
\begin{table}[htbp!]
    \caption{Lattice parameters ($a_0$) of two-dimensional systems studied in this work.}
\label{tab:mono_lattice}
\renewcommand{\arraystretch}{1.2} 
\tablinesep=5ex\tabcolsep=10pt
\begin{tabular}{lcc}  
\toprule
          & $a_0(\AA)$       &  Reference \\
\midrule
  $\mathrm{BP}$              &  3.18       & Ref.~\citenum{PhysRevB.80.155453}       \\
  $\mathrm{phosphorene}$     &  3.31       & Ref.~\citenum{PhysRevB.89.235319}      \\
  $\mathrm{MoS_2}$           &  3.16       & Ref.~\citenum{huser2013dielectric}     \\
  $\mathrm{WS_2}$            &  3.15       & Ref.~\citenum{chen2022nonunique}      \\
  $\mathrm{GaAs}$            &  3.97       & Ref.~\citenum{fakhrabad2014quasiparticle}     \\
  $\mathrm{GaN}$             &  3.20       & Ref.~\citenum{SHU2019475}     \\
  $\mathrm{AlN}$             &  3.09       & Ref.~\citenum{PhysRevB.80.155453}      \\
  $\mathrm{graphane (CH)}$   &  2.51       & Ref.~\citenum{PhysRevMaterials.2.124002}     \\
   h-BN                      &  2.51       & Ref.~\citenum{PhysRevB.87.035404}      \\
   \bottomrule
\end{tabular}
\renewcommand{\arraystretch}{1} 
\end{table}

\section{Electronic structure of complex materials using the SE-RSH functional\label{sec3}}

In the following, we first assess the accuracy of the SE-RSH functional for the electronic properties of three-dimensional homogeneous systems and then investigate heterogeneous systems, including interfaces, two-dimensional pristine and defective systems (2D systems) and nanoparticles. 

\subsection{Homogenous 3D systems}
\begin{figure}[htbp!]
\includegraphics[width=8.5cm]{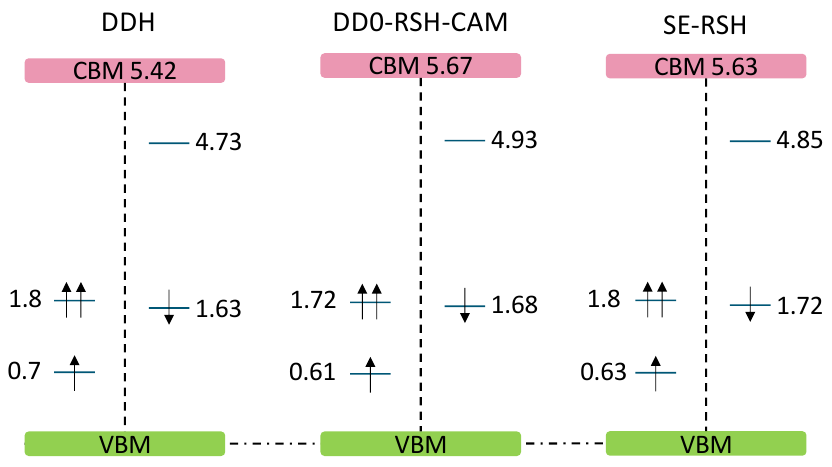}
\caption{Single-particle energy levels (eV) of the nitrogen vacancy in diamond, computed at different levels of theory and referenced to the valence band maximum (VBM). We show results obtained with the DDH\cite{PhysRevB.89.195112}, DD0-RSH-CAM\cite{10.1103/physrevmaterials.2.073803} functionals and the SE-RSH functional proposed in this work. All calculations were performed in cells with 216 atoms.} \label{pic:nv_diamond}
\end{figure}

In Table \ref{tab:3Dsystems}, we compare the band gaps obtained for 3D systems using the SE-RSH hybrid functional and other dielectric-dependent hybrid functionals\cite{PhysRevB.89.195112, 10.1103/physrevb.93.235106, 10.1103/physrevmaterials.2.073803}, specifically the RS-DDH\cite{10.1103/physrevb.93.235106} and the DD0-RSH-CAM\cite{10.1103/physrevmaterials.2.073803}, that both utilize $1/\epsilon_{\infty}$ as the fraction of Fock exchange in the long range part and a constant screening parameter $\mu$. The main difference between RS-DDH and DD0-RSH-CAM lies in the description of the short-range exchange potential: the fraction of Fock exchange is 25\% in the former and 100\% in the latter. In our comparisons, the average values of the local dielectric function $\overline{\epsilon(\mathbf{r})}$ and of the local screening function $\overline{\mu(\mathbf{r})}$ are adopted when using RS-DDH and DD0-RSH-CAM, instead of $\epsilon_\infty$ and $\mu$.\\
\indent Since in our calculations we did not consider electron-phonon coupling, we subtracted the value of the zero-phonon remormalization (ZPR)\cite{10.1103/physrevmaterials.2.073803} from the computed values of the band gaps of semiconductors and insulators when comparing with experiments.  
\begin{table*}[htbp!]
    \caption{Fundamental energy gaps (eV) [Columns 4-8] of three dimensional materials obtained  using different energy functionals, as specified in the first row. Column 2 and 3 provide the macroscopic dielectric constant and constant screening parameter used in DDH, RS-DDH and DD0-RSH-CAM calculations. Zero-phonon renormalization of the band gap ($E_{\rm{corr}}$) (eV) are also reported in the last column when available from experiments. MAE and MARE are the mean absolute and mean absolute relative error of calculated band gap, respectively, compared to ($\mathrm{Exp.} + E_{\mathrm{corr}}$).}
\label{tab:3Dsystems}
\begin{threeparttable}
\renewcommand{\arraystretch}{1.2} 
\tabcolsep=2.7pt 
\begin{tabular}{l | c c | c c c c c c c c }
\toprule
             & {\footnotesize $\overline{\epsilon(\mathbf{r})}$ }& {\footnotesize$\overline{\mu(\mathbf{r})}$} & {\footnotesize PBE\cite{10.1103/physrevlett.77.3865}}   & {\footnotesize HSE06\cite{10.1063/1.1564060, 10.1063/1.2204597}} & {\footnotesize DDH\cite{PhysRevB.89.195112}} & {\footnotesize RS-DDH\cite{10.1103/physrevb.93.235106}} & {\footnotesize DD0-RSH-CAM\cite{10.1103/physrevmaterials.2.073803}} & {\footnotesize SE-RSH} &  {\footnotesize Exp.} & {\footnotesize $E_{\rm{corr}}$\tnote{a}} \\
\midrule
    Si            & 11.76 & 0.55 & 0.61 & 1.14 & 0.99 &   1.02  & 1.11   &  1.10  &  1.17 & 0.06              \\
    $\rm{SiO_2}$  & 2.49 & 0.51 &  5.95 & 7.69 & 9.99 &   9.88  & 10.62  &  10.43 &  9.7              \\
    C             & 5.60 & 0.68 &  4.19 & 5.35 & 5.42 &   5.45  & 5.67   &  5.63  &  5.48 & 0.37             \\
    SiC           & 6.53 & 0.61 &  1.37 & 2.24 & 2.35 &   2.36  & 2.45   &  2.44  &  2.42 & 0.11             \\
    BN            & 4.43 & 0.68 &  4.53 & 5.83 & 6.33 &   6.39  & 6.56   &  6.56  &  6.4 & 0.34            \\
    AlP           & 7.27 & 0.55 &  1.56 & 2.25 & 2.27 &   2.37  & 2.47   &  2.45  &  2.52 & 0.02            \\
    AlAs          & 8.27 & 0.57 &  1.32 & 1.93 & 1.91 &   1.90  & 2.07   &  2.05  &  2.24 & 0.04             \\
    GaAs          &13.37 & 0.54 &  0.42 & 1.26 & 0.79 &   0.90  & 1.43   &  1.41  &  1.52 & 0.05             \\
    MgO           & 2.81 & 0.63 &  4.79 & 6.47 & 7.70 &   7.88  & 8.32   &  8.25  &  7.83 & 0.53             \\
    \midrule
     {\footnotesize MAE(eV)}      &     & &  1.78    & 0.74 &  0.40    &  0.33     &  0.22    &    0.22     &      &        \\
     {\footnotesize MARE(\%)}      &    & &  43.6    & 14.4 &  14.1    &  12.16     & 5.5     &   5.9      &      &         \\
\bottomrule
\end{tabular}
\renewcommand{\arraystretch}{1} 
\begin{tablenotes}
\item [$^a$]$E_{\rm{corr}}$ values are from Ref.~\citenum{10.1103/physrevmaterials.2.073803}.
\end{tablenotes}
\end{threeparttable}
\end{table*}
 The results obtained with the SE-RSH and DD0-RSH-CAM functionals are extremely similar, showing the relative insensitivity of the band gaps of the semiconductors studied here to the small variation of $\epsilon(\mathbf{r})$ and $\mu(\mathbf{r})$ in homogeneous 3D systems\cite{10.1103/physrevmaterials.3.073803}. We note that when $\epsilon(\mathbf{r})$ and $\mu(\mathbf{r})$ are constant, the expressions of the SE-RSH and DD0-RSH-CAM functionals coincide. The energy gaps calculated using the RS-DDH functional are marginally lower compared to those obtained with the DD0-RSH-CAM and SE-RSH functionals, indicating that using the unscreened Fock exchange potential within the short-range component is an effective strategy.\\
\indent We used the same functionals to compute the electronic structure of point defects in 3D materials, in particular the nitrogen-vacancy (NV) center in diamond. The results are presented in Fig.(\ref{pic:nv_diamond}). We find that SE-RSH yields band edges and defect levels comparable to those obtained from DD0-RSH-CAM calculations, showing that using the average local dielectric function and local screening parameter of the host crystal provides an accurate description of the electronic properties of the defect. We will see in Sec. \ref{sec2:defective2D} that a different conclusion holds for defects in 2D systems.   

\subsection{Heterogeneous 3D systems: Interfaces}

We investigated the electronic properties of several representative interfaces H-$\mathrm{Si/H_2O}$, Si/$\rm{Si_3N_4}$, Si/$\rm{SiO_2}$) using the same models reported in Ref.~\citenum{10.1103/physrevmaterials.3.073803, 10.1021/acs.jctc.7b00853}.

\begin{figure*}[htbp!]
\includegraphics[width=16cm]{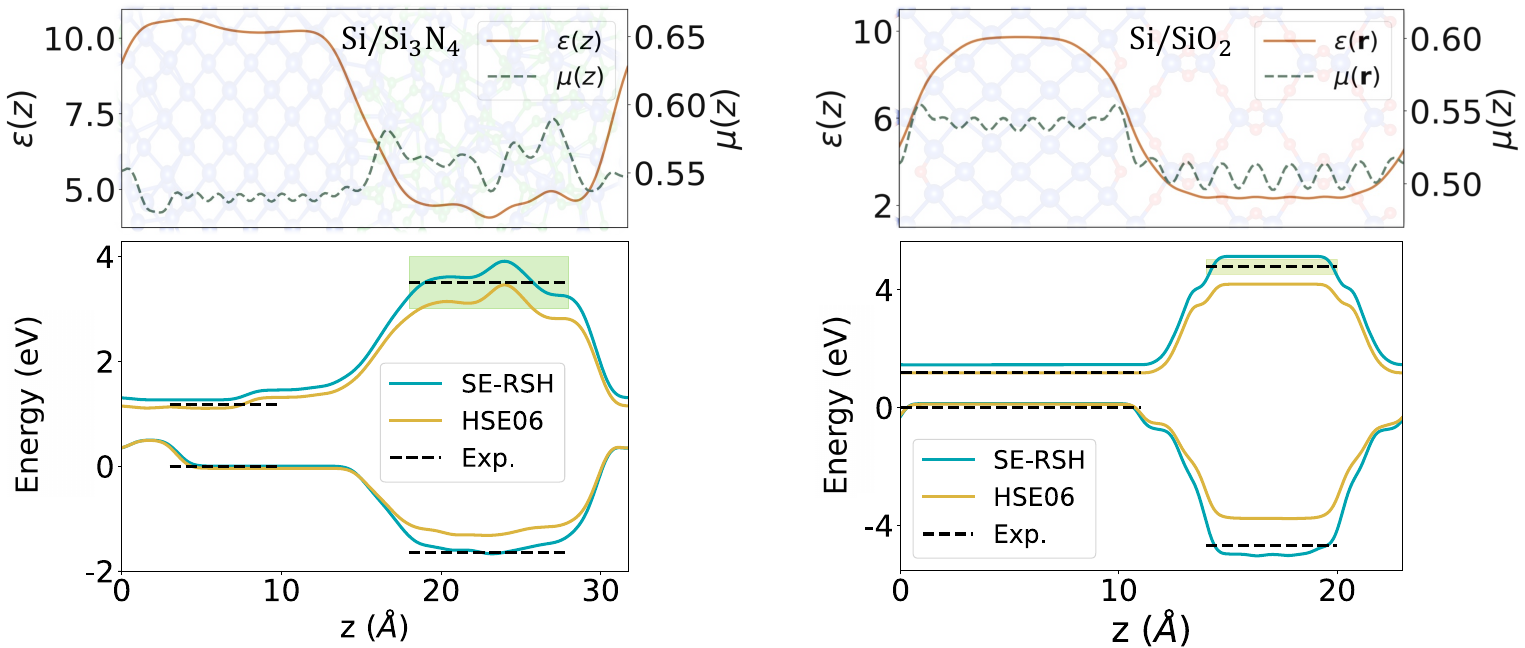}
\caption{Top panel: The local dielectric function $\epsilon(z)$ and local screening function $\mu(z)$ for model Si/$\rm{Si_3N_4}$ and Si/$\mathrm{SiO_2}$ interfaces, computed as averages of $\epsilon(\mathbf{r})$ (Eq. \ref{epsilon}) and $\mu(\mathbf{r})$ (Eq. \ref{mu}) in the (x,y) plane parallel to the interface. The z axis is perpendicular to the interface. Bottom panel: Band edges of Si/$\rm{Si_3N_4}$ and Si/$\mathrm{SiO_2}$ computed (Eq. \ref{ldos}) at different levels of theory, using the SE-RSH functional proposed in this work and the HSE06 functional\cite{10.1063/1.1564060, 10.1063/1.2204597}. Experimental data (Exp.)\cite{10.1063/1.4811481, bersch2010complete} are reported as dashed lines and green bands, whose width indicates the expected range of the conduction band minimum (See Table \ref{tab:1}).} \label{pic:interface}
\end{figure*}

\indent Fig.(\ref{fig:1}) and (\ref{pic:interface}) show the band edges of three interfaces obtained by computing the Local Density of States (LDOS)\cite{yamasaki2001geometric} defined as:
\begin{equation}
    D(\epsilon, z) = 2 \sum_{n}|\langle z|\phi_n\rangle|^2\delta(\epsilon - \epsilon_n), \label{ldos}
\end{equation}
where z is the direction perpendicular to the interface, $|\langle z|\phi_n\rangle|^2$ is the electron density integrated in the $xy$ plane, and $\epsilon_{n}$ are the Kohn-Sham eigenvalues.\\
\indent In the case of H-Si/$\mathrm{H_2O}$, we obtain a band gap in the silicon region (1.8 eV) which is larger than the bulk value (1.17 eV) since the slab chosen to represent Si has just few layers and hence the system is quantum confined\cite{weston2018accurate}. In the water region, we obtain a value of 10 eV, which is in good agreement with the values 10.5 eV obtained at the $\mathrm{G_0W_0@}$DDH level\cite{gaiduk2018electron}. Interestingly, SE-RSH provides a smaller band gap compared to DD0-RSH-CAM, which yields 10.9 eV\cite{10.1103/physrevresearch.3.023182}. This discrepancy can be attributed to the more localized nature of the electronic density in water, relative to that of semiconductors reported in Table \ref{tab:3Dsystems}. In water, $\epsilon(\mathbf{r})$ exhibits marked fluctuations, reaching values substantially higher than $\epsilon_\infty$ in regions of high electronic density. Hence, using $1/\epsilon_\infty$ as the mixing fraction of exact exchange in the long-range component is not an accurate approximation. In the cases of Si/$\rm{Si_3N_4}$ and Si/$\rm{SiO_2}$, we also compare computed band offsets with experiments and other calculations (see Table \ref{tab:1}) carried out using supercells of the same size.

\begingroup
\setlength{\tabcolsep}{10pt} 
\renewcommand{\arraystretch}{1.2} 

\begin{table} [htbp!]
    \caption{Conduction band (CBO) and Valence band (VBO) offsets (eV) for models of $\mathrm{Si/Si_3N_4}$ and $\mathrm{Si/SiO_2}$ interfaces computed at different levels of theory, compared to experimental results (Exp.). The $G_0W_0$ results have been computed starting from PBE calculations.} 
    \label{tab:1}
  \begin{threeparttable}
     \begin{tabular}{llcc}
        \toprule
        Interface & Method  & CBO & VBO \\
        \midrule
  \multirow{4}{*}{$\mathrm{Si/Si_3N_4}$}      & HSE06       & 1.90       & 1.23   \\
                                              & SE-RSH     & 2.3        &  1.63    \\
                                              & $G_0W_0$\cite{10.1063/1.4811481}     & 1.90 & 1.50\\
                                              & Exp.\cite{10.1063/1.4811481}        & 1.83-2.83  & 1.5-1.78    \\
                                              &             &            &    \\
   \multirow{4}{*}{$\mathrm{Si/SiO_2}$}       & HSE06       &  3.02      & 3.75    \\
                                              & SE-RSH     & 3.68       &  5.07     \\
                        & $G_0W_0$\cite{10.1021/acs.jctc.7b00853} & 2.90 & 4.10 \\
                                              & Exp.        & 3.22-3.82$^{a}$       & 4.71\cite{bersch2010complete}    \\
        \bottomrule
     \end{tabular}
    \begin{tablenotes}
      \small
      \item $^a$CBO estimated using VBO=4.71 eV, and the experimental band gaps of $\rm{Si}$ (1.17 eV) and $\rm{SiO_2}$ (9.1–9.7 eV\cite{10.1021/acs.jctc.7b00853, keister1999band}), respectively.
    \end{tablenotes}
  \end{threeparttable}
\end{table}
\endgroup

 We now turn to semiconductor superlattices that are almost lattice-matched, and represent a class of systems  thoroughly studied\cite{steiner2014band} by the electronic structure community. We computed band offsets for $\mathrm{GaAs/AlAs(100)}$, $\mathrm{AlP/GaP(100)}$ and $\mathrm{Si/GaP(100)}$ heterojunctions, using the same supercells as in Ref.~\citenum{steiner2014band} and the method of Van de Walle et al.\cite{van1987theoretical}. Our results are reported  in Table \ref{tab:2}, together with those of other calculations performed with the PBE, PBE0 functionals and the $G_0W_0$@PBE approximation\cite{10.1103/physrevmaterials.3.073803, 10.1021/acs.jctc.7b00853}.

\begingroup
\setlength{\tabcolsep}{10pt} 
\renewcommand{\arraystretch}{1.2} 

\begin{table} [htbp!]
\caption{Conduction (CBO) and valence (VBO) band offsets of selected semiconductor heterojunctions computed using the method of Ref.~\citenum{van1987theoretical}.}
\label{tab:2}
\begin{threeparttable}
     \begin{tabular}{llcc}
        \toprule
         & Method$^{a}$  & CBO & VBO \\
        \midrule
   \multirow{5}{*}{$\mathrm{AlAs/GaAs}$}      & HSE06       & 0.26      & 0.43    \\
                                              & SE-DDH        & 0.57          & 0.54  \\
                                              & SE-RSH     & 0.11      & 0.56     \\
                                              & $G_0W_0$ & 0.17  & 0.60    \\
                                              & Exp.        & 0.18       & 0.53    \\
                                              &             &            &    \\
   \multirow{5}{*}{$\mathrm{GaP/Si}$}         & HSE06       & 0.74       & 0.40   \\
                                              & SE-DDH        & 0.72       & 0.47   \\
                                              & SE-RSH     & 0.71       & 0.55    \\
                                              & $G_0W_0$ & 0.83  & 0.53    \\
                                              & Exp.        & 0.38       & 0.80    \\
                                              &             &            &    \\
   \multirow{5}{*}{$\mathrm{AlP/GaP}$}        & HSE06       & -0.55      & 0.48    \\
                                              & SE-DDH        & -0.40      & 0.56   \\
                                              & SE-RSH     & -0.50      & 0.55     \\
                                              & $G_0W_0$ & -0.87  & 0.67    \\
                                              & Exp.        & -0.39       & 0.55    \\
        \bottomrule
     \end{tabular}
     \begin{tablenotes}
      \small
      \item $^a$ Experimental and $G_0W_0$ values are from Ref.~\citenum{10.1021/acs.jctc.7b00853}.
    \end{tablenotes}
     \end{threeparttable}
\end{table}
\endgroup

In agreement with $G_0W_0$ calculations, all hybrid functionals show that $\mathrm{AlAs/GaAs}$ and $\mathrm{Si/GaP}$ are type \uppercase\expandafter{\romannumeral 1\relax} heterojunctions, whilst $\mathrm{AlP/GaP}$ is type \uppercase\expandafter{\romannumeral 2\relax}. The SE-DDH functional, however, yields a larger value for the conduction band offset (CBO) of AlAs/GaAs due to its substantial underestimation of the experimental GaAs band gap. Indeed, hybrid functionals without a correct short-range mixing fraction commonly underestimate the band gap of systems with localized semicore d states\cite{10.1103/physrevmaterials.2.073803}. Instead, the accuracy of the SE-RSH functional in predicting band offsets is comparable to that of $G_0W_0$, due to the description, on the same footing, of localized and delocalized electronic states.

\subsection{2D systems\label{sec: 2d}}

Predicting the electronic structure of 2D systems is more challenging, in general, than that of 3D materials, as many xc functionals are tailored for bulk solids\cite{PhysRevLett.102.226401}. For example, the DDH functional is not expected to be accurate for 2D systems because the macroscopic dielectric constant $\epsilon_{\infty}$ is ill-defined when vacuum is present in the supercell. In addition the comparison with experiments is not straightforward since measurements are usually performed on substrates whose interaction with the 2D material may not be negligible\cite{10.1038/srep29184}.
\begingroup
\setlength{\tabcolsep}{8pt} 
\renewcommand{\arraystretch}{1.2} 
\begin{table*} [htbp!]
\caption{ Fundamental energy gaps (eV) evaluated with hybrid functionals compared with the results of  $GW$ calculations for  2D systems. For calculations with the DDH functional we used the mixing fraction corresponding to the 3D bulk long-wavelength dielectric constant. The SE-DDH and SE-RSH columns report electronic gaps evaluated with the functionals described in Sec.\ref{sec1:level2}. The calculations marked as Ref. ($GW$) have all been performed at the $G_0W_0$@PBE level of theory (except those of Ref.~\citenum{PhysRevB.80.155453}\tnote{a}) with the lattice constant shown in Table \ref{tab:mono_lattice}. The MAE and MARE are computed by comparing the calculated band gap to the reference gap (column 6).
} \label{tab:monolayer}
  \begin{threeparttable}
\renewcommand{\arraystretch}{1.2} 
\tabcolsep=2.5pt
     \begin{tabular}{lcccccc}
        \toprule
                    & HSE06\cite{10.1063/1.1564060, 10.1063/1.2204597}      & DDH\cite{PhysRevB.89.195112} & SE-DDH\cite{10.1103/physrevmaterials.3.073803}     & SE-RSH        & Ref. ($GW$) & Other $GW$ results\tnote{b} \\
        \midrule
$\mathrm{BP}$              &  1.36       & 1.35                             &  1.31    & 1.64          &  1.81\tnote{a}$\ \ $ \cite{PhysRevB.80.155453}       \\
   $\mathrm{phosphorene}$     &  1.50       & 1.60                             &  1.66    &  2.1         &  2.00  \cite{PhysRevB.89.235319} & 1.94-2.1\cite{liang2014electronic, chen2018tunable}      \\
   $\mathrm{MoS_2}$\tnote{c}           &  2.17       & 2.3                            &  2.13    & 2.61          &  2.58   \cite{huser2013dielectric} & 2.60-2.80\cite{cheiwchanchamnangij2012quasiparticle, shi2013quasiparticle, jin2015tuning}     \\
   $\mathrm{WS_2}$\tnote{c}            &  2.48        & 2.46                            &  2.50    & 2.68           &  2.90   \cite{chen2022nonunique} & 3.05-3.11\cite{shi2013quasiparticle,wang2014many}      \\
   $\mathrm{GaAs}$            &  1.67       & 1.63                           &  1.63    & 2.43           &  2.86   \cite{fakhrabad2014quasiparticle} & 2.95\cite{mishra2020exciton}     \\
   $\mathrm{GaN}$             &  3.53       & 3.64                            &  3.97    & 4.41           &  4.44   \cite{SHU2019475} & 4.14-5.00\cite{chen2011tailoring, PhysRevB.80.155453}     \\
   $\mathrm{AlN}$\tnote{c}             &  3.08       & 4.73                           &  5.33    & 5.68         &  5.57\tnote{a}$\ \ $      \cite{PhysRevB.80.155453} & 5.36-5.7\cite{paperid:1050139, karami2021tuning}      \\
   $\mathrm{graphane (CH)}$   &  4.41       &                                      &  5.40    & 6.02           &  6.40    \cite{PhysRevMaterials.2.124002}     \\
    h-BN                      &  5.7        & 6.41                           &  7.27    & 7.52           &  7.40\   \cite{PhysRevB.87.035404}  & 7.00-8.43\cite{attaccalite2011coupling, huser2013quasiparticle, PhysRevB.87.035404, roman2021band}     \\
    \hline
     MAE(eV)                   &  1.28       & 0.79                                   &  0.58    & 0.18\\
     MARE(\%)                  &  29.4      & 21.1                                    &  17.9   & 5.4\\
        \bottomrule
     \end{tabular}
     \renewcommand{\arraystretch}{1} 
\begin{tablenotes}
\item [$^a$]$G_0W_0$@LDA;
\item [$^b$]Results obtained with lattice parameters different from those of Table \ref{tab:mono_lattice} or $GW$ approximation different from $G_0W_0\mathrm{@PBE}$;
\item [$^c$]Kinetic energy cutoff: 70 Ry
\end{tablenotes}
  \end{threeparttable}
\end{table*}
\endgroup
 \\ \indent Here we assess the accuracy of the SE-RSH functional for nine 2D systems by comparing results with those of the $GW$ quasiparticle method. Table \ref{tab:monolayer} shows the results obtained with several hybrid functionals, including SE-RSH, SE-DDH (Eq. \ref{equ: 5}), HSE06 and DDH ($\alpha=1/\epsilon_{\mathrm{bulk}}$). Values of the macroscopic dielectric constant for their corresponding bulk systems ($\epsilon_{\mathrm{bulk}}$) used in DDH calculations are from Ref.~\citenum{10.1103/physrevmaterials.2.073803}.  $GW$ quasiparticle band gaps are used as reference values. We find that HSE06 systematically underestimates the fundamental band gaps of all the 2D systems studied here and that 
 DDH yields more accurate band gaps compared to HSE06, albeit not in excellent agreement with experiments. The accuracy of DDH for 2D systems may be improved\cite{PhysRevMaterials.2.124002} if the mixing fraction of exact exchange is appropriately fine-tuned through the enforcement of the generalized Koopmans’ condition\cite{10.1103/physrevlett.49.1691, perdew2017understanding}, e.g. by using specific defects as probes. However, such a method may require multiple calculations for charged defective systems whose specific choice is material dependent. \\
\indent When using SE-DDH, we obtain an improvement over DDH results, however the computed  band gaps are slightly underestimated compared to our reference results.\\
\begin{figure}[htbp!]
\includegraphics[width=8.6cm]{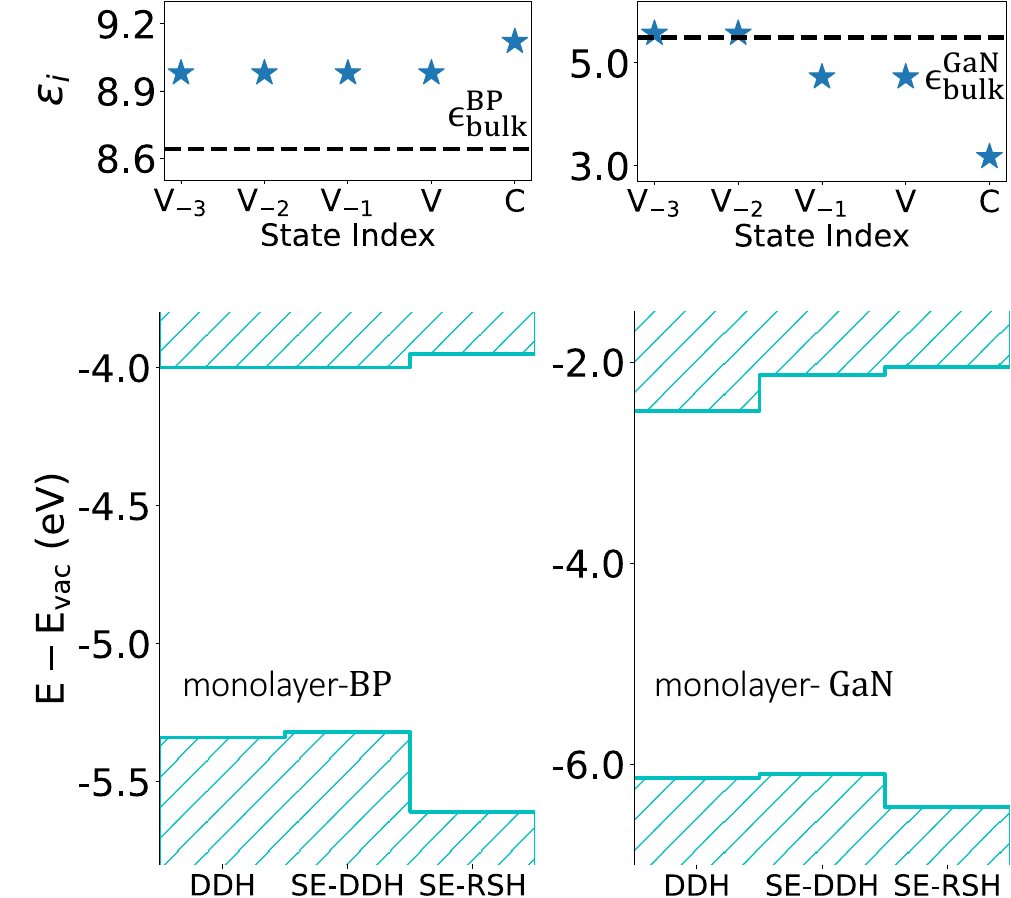}
\caption{\label{fig:BP_GaN} Upper panel: The state-dependent screening $\epsilon_i$ (Eq. \ref{eq: epsilon_i}) as a function of the state index for selected bands (V and C denote VBM and CBM respectively) for monolayer BP (left) and GaN (right). The dashed lines show the values of the long-wavelength dielectric constant of the 3D bulk ($\epsilon_{\mathrm{bulk}}$). Bottom panel: Single-particle eigenvalues referenced to vacuum computed with the DDH\cite{PhysRevB.89.195112} and SE-DDH\cite{10.1103/physrevmaterials.3.073803} functionals, and the SE-RSH functional proposed in this work. Calculations with the DDH functional were performed with $\alpha=1/\epsilon_{\mathrm{bulk}}$.}
\end{figure}
\indent Among all hybrid functionals adopted here, the SE-RSH functional provides the most accurate band gaps. It usually yields larger band gaps than SE-DDH, aligning more closely with the $GW$ results. This effect is particularly pronounced for ML-GaAs, illustrating that the underestimation of band gaps in systems with localized semicore d states by SE-DDH, previously observed in semiconductor heterojunctions, is also evident in two-dimensional systems.\\
\indent  Although Ref.~\citenum{PhysRevMaterials.2.124002, chen2022nonunique} mainly attribute the failure of  DDH to the inability to describe the weak screening of the vacuum surrounding  2D systems, we find that in some cases simply using a local dielectric function $\epsilon(\mathbf{r})$ does not improve the band gaps obtained with DDH. For example, in monolayer BP system, both SE-DDH and DDH predicts a band gap $\simeq$ 1.3 eV, while the $G_0W_0$@PBE value is 1.8 eV. 

To understand these results, we further consider the electronic structures of two representative monolayers: BP and GaN, and  we define state-dependent screening parameters as
\begin{equation}
    \epsilon_i = \int\epsilon(\mathbf{r})|\phi_i(\mathbf{r})|^2\mathrm{d}\mathbf{r} \label{eq: epsilon_i}
\end{equation}
to obtain an estimate of the effect of screening on each band.\\
\indent Fig.(\ref{fig:BP_GaN}) shows the variation of computed band edges as a function of the exchange-correlation functionals, along with the state-dependent screenings of several bands (VBM and CBM are the highest occupied and lowest unoccupied bands, respectively). For monolayer BP, the screening of both VBM and CBM is close to $\epsilon_{\mathrm{bulk}}^{\mathrm{BP}}$, and SE-DDH and DDH yield similar results. However, for monolayer GaN,  the screening of VBM and CBM are smaller than $\epsilon_{\mathrm{bulk}}^{\mathrm{GaN}}$ and  SE-DDH yields higher CBM and VBM than DDH and also a larger band gap. For both BP and GaAs,  SE-RSH further increases the band gap and yields more accurate results.\\
\indent The above examples illustrate that in certain low-dimensional systems, electronic states may be subject to bulk-like screening when the in-plane macroscopic dielectric constant of the monolayerr is similar to $\epsilon_{\infty}$ of the 3D bulk\cite{laturia2018dielectric}. In this case, it is the inaccurate description of short-range screening that may lead to the failure of DDH.

\subsection{Defects in 2D systems\label{sec2:defective2D}}
 \indent In the literature, several first-principle calculations of the electronic structure of defects in 3D systems have been performed using a global hybrid functional, known as PBE0($\alpha$)\cite{miceli2018nonempirical}, where the amount of Fock exchange admixed with semilocal exchange is controlled by a single parameter $\alpha$; the latter is determined by fitting the experimental value of the band gap of the host system.\\
 \indent However, such a scheme often fails for 2D systems\cite{chen2022nonunique}. Here, we investigate various point defects in monolayer systems, including a single carbon atom substituting sulfur ($\rm{C_S}$) in monolayer $\mathrm{WS_2}$ and substituting boron ($\rm{C_B}$) in monolayer h-BN.\\
 \begin{figure}[htbp!]
\includegraphics[width=8.6cm]{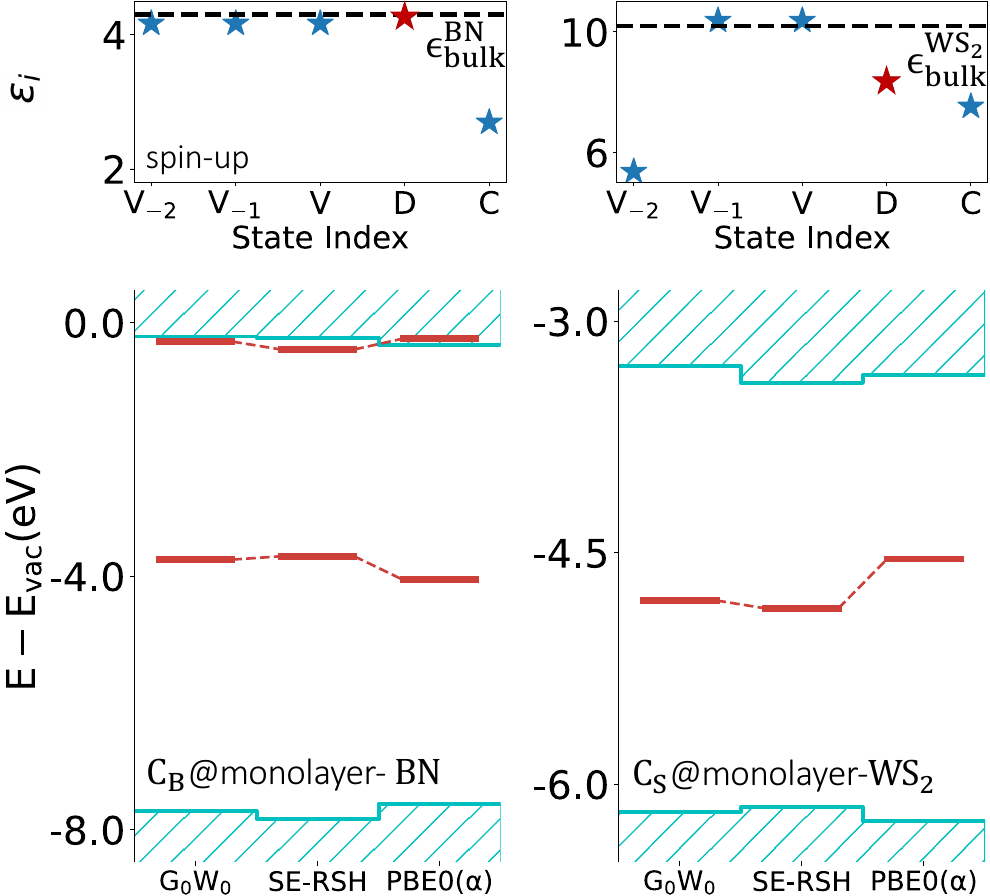}
\caption{\label{fig:WS2-HBN}Upper panel: The state-dependent screening $\epsilon_i$ (Eq. \ref{eq: epsilon_i}) is plotted as a function of state index (V, D and C denote VBM, defect state and CBM respectively). The dashed lines represent the macroscopic dielectric constant of the corresponding bulk systems. Lower panel: Single-particle eigenvalues with respect to the vacuum level. For each system, SE-RSH and PBE0($\alpha$) results are compared against $G_0W_0$ reference results.}
\end{figure}
 \indent Fig.(\ref{fig:WS2-HBN}) shows the computed single-particle eigenvalues referenced to vacuum obtained with different electronic structure methods, including ${G_0W_0}$@PBE. Also showed in Fig.(\ref{fig:WS2-HBN}) is the state-dependent screening defined in Eq.(\ref{eq: epsilon_i}).\\
\indent We found that for the two defective systems studied here, PBE0($\alpha$) predicts qualitatively inaccurate results compared to $GW$. In the case of ML h-BN, for example, PBE0($\alpha$) would incorrectly predict one defect state to be above the CBM.  This result may stem from the differences in dielectric screening experienced by the defect states and the band edges of the material. As illustrated in the upper panel of Fig.(\ref{fig:WS2-HBN}), the CBM of ML h-BN is exposed to a weaker screening than the defect states and the VBM, and hence a mixing fraction determined solely from the band gap of ML h-BN is likely too large to accurately describe defect levels.\\
\indent In the case of ML $\rm{WS2}$, the discrepancy between PBE0($\alpha$) and SE-RSH or $GW$ results arises not only from the difference in dielectric screening experienced by different electronic states, but also from the difference in the level of orbital localization, in particular the d orbitals of W contributing to the VBM and CBM. 

\indent Unlike PBE0,  SE-RSH can effectively differentiate between diverse electronic states based on their respective environments and underlying physics and hence yields more accurate results.

\subsection{Finite Systems}
\begin{figure}[htbp!]
\includegraphics[width=8.6cm]{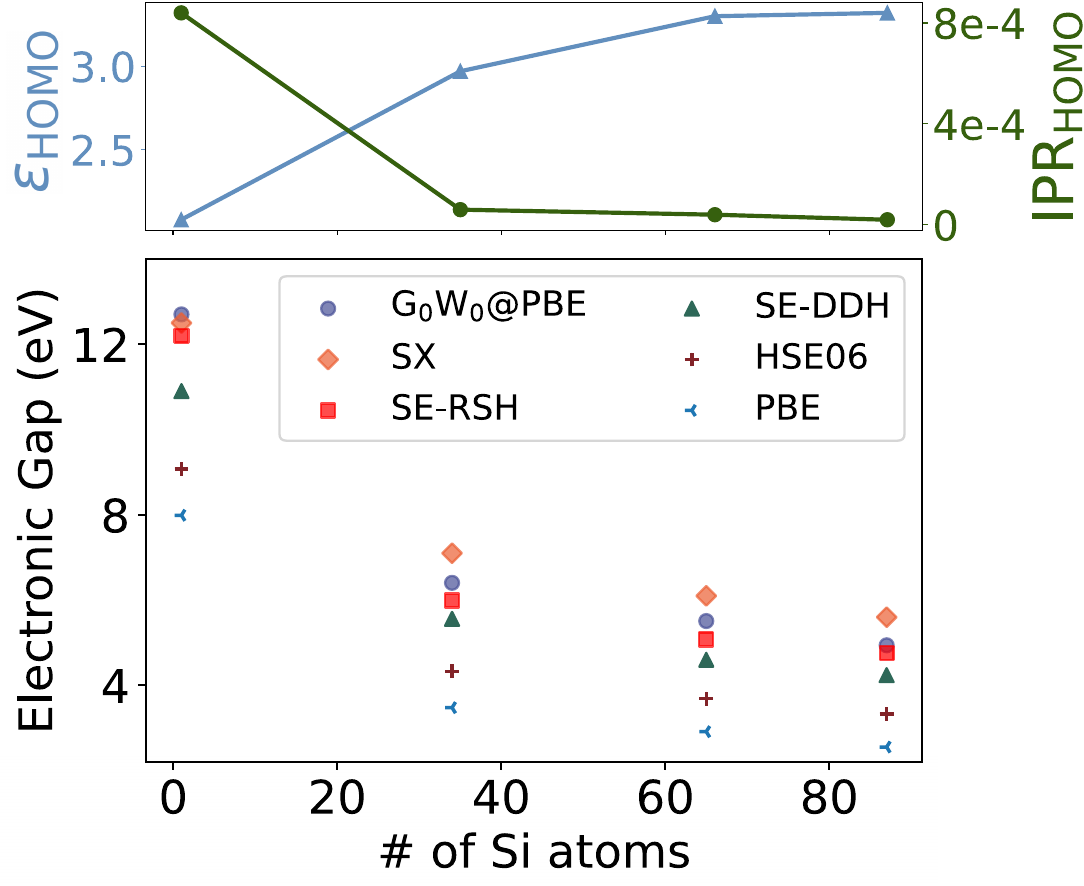}
\caption{\label{fig:4} Upper panel: The screening (Eq. \ref{eq: epsilon_i}) and inverse participation ratio (IPR) of the highest occupied molecular orbital (HOMO) as a function of the number of Si atoms in the core of hydrogen-passivated silicon clusters. Bottom panel: The fundamental gaps of hydrogen-passivated silicon clusters as a function of the number of Si atoms in the core, predicted using SE-RSH, SE-DDH\cite{10.1103/physrevmaterials.3.073803}, SX\cite{10.1021/acs.jctc.7b00368} and HSE06\cite{10.1063/1.1564060, 10.1063/1.2204597}. Fundamental gaps obtained from $G_0W_0$ starting from PBE wave functions ($G_0W_0$@PBE) are included for comparison. }
\end{figure}
Similar to the case of low dimensional systems, DDH is not expected to be accurate for finite systems such as nanoparticles. A generalization of DDH to finite systems was proposed by Brawand et al.\cite{10.1021/acs.jctc.7b00368} (we call this generalization SX), with:
\begin{equation}
    v_{xc}^{\rm{SX}}(\mathbf{r, r'}) = \alpha^{\mathrm{SX}}\Sigma_{x}(\mathbf{r, r'}) + (1-\alpha^{\mathrm{SX}})v_{x}(\mathbf{r}) + v_{c}(\mathbf{r}),
\end{equation}
where $\alpha^{\mathrm{SX}}$ is the screened-exchange constant:
\begin{equation}
    \alpha^{\mathrm{SX}} = \frac{\sum_{j}^{N}\langle j | \Sigma_{\mathrm{SEX}} | j \rangle}{\sum_{j}^{N}\langle j | \Sigma_x | j \rangle}.
\end{equation}
The SX functional is, in spirit, similar to SE-RSH since the mixing fraction $\alpha^{\mathrm{SX}}$ is also designed to replace the non-local exchange potential with the screened-exchange self-energy $\Sigma_{\mathrm{SEX}}$. However it is less general than SE-RSH. 
\\ \indent Fig.(\ref{fig:4}) presents the fundamental gaps of several hydrogen-passivated silicon clusters, computed using various methods and plotted against the number of Si atoms within the cluster core, using the  structural models of Ref.~\citenum{article}. The Mean Absolute Error (MAE) of SE-RSH with respect to $G_0W_0$@PBE is found to be approximately 0.19 eV, as compared to  0.27, 0.53 and 1.14 eV for SX, SE-DDH and HSE06, respectively.  The PBE functional markedly underestimates the gap of these clusters, with a MAE of 1.58 eV. Interestingly,  the fundamental gaps predicted by HSE are consistently shifted by about 0.8 eV relative to the corresponding PBE values, suggesting a similar description of quantum confinement effects by the two functionals, albeit distinct from that obtained from the $GW$ approximation.\\
\indent In Fig.(\ref{fig:4}) we also present the state-dependent screening values as well as the inverse participation ratio of the highest occupied orbitals, defined as $\mathrm{IPR}_{\mathrm{HOMO}}=\int|\phi_{\mathrm{HOMO}}(\mathbf{r})|^4\mathrm{d}\mathbf{r}$. A higher IPR value indicates a more localized single particle
wave function. We can see that as the size of the system is reduced, $\epsilon_{\mathrm{HOMO}}$ decreases. Such a change leads to a size-dependent correction, relative to PBE, obtained with both SE-DDH and SE-RSH. However, SE-DDH becomes more inaccurate as the IPR increases, while SE-RSH provides a consistently accurate description of the gaps for all molecules, demonstrating the accuracy of SE-RSH in dealing with strongly localized electronic systems and the importance of enhanced short-range Fock exchange for finite systems.
 
\section{Conclusions\label{sec:4}}
In this work, we generalized the definition of range-separated hybrid functionals to heterogeneous systems by defining  a non-empirical functional (SE-RSH), where the mixing fraction of exact and local exchange depends on a spatially dependent local dielectric function and a local screening function. The definition of SE-RSH is inspired by the static COHSEX approximation\cite{PhysRevLett.55.1418}, used in many body perturbation theory. Using the  SE-RSH, we obtained electronic energy gaps and band offsets in good agreement with experimental data for both homogeneous and heterogeneous three-dimensional systems, and in excellent agreement with $GW$ calculations for pristine and defective two-dimensional systems as well as silicon clusters for which measurements are currently unavailable. This good agreement stems from the ability of the SE-RSH functional to account for the different screening experienced by different electronic states and by an accurate description  of both localized and delocalized electronic states. Given its accuracy and its derivation without any empirical parameters, the SE-RSH functional is particularly well-suited for high-throughput calculations of heterogeneous systems. Additionally, research is currently underway to explore the integration of the SE-RSH functional with time-dependent DFT (TDDFT)\cite{runge1984density, jin2022vibrationally} calculations, thus enabling efficient investigations of excited-state properties and optical response in complex materials.

\begin{suppinfo}

Details of implementation, including the procedure to obtain a coarse-grained approximation of the local screening function $\mu(\mathbf{r})$ as well as the method to handle the integrable divergence of the exchange energy.

\end{suppinfo}

\begin{acknowledgement}

We thank Francois Gygi, Yu Jin and Huihuo Zheng for many useful discussions. This work was supported by DOE/BES through the computational materials science center Midwest Integrated Center for Computational Materials (MICCoM). The computational resources were provided by University of Chicago’s Research Computing Center. 

\end{acknowledgement}

\bibliography{se-rsh}


\end{document}


\section{Coarse-Grained Approximation of $\mu(\mathbf{r})$\label{cga}}
In SE-RSH, the exchange energy $E_{x, i}$ associated to the state $\phi_i$ is:
\begin{equation}
    E_{x, i} = \sum_j \iint \frac{\alpha(\mathbf{r, r'})\rho_{ij}^*(\mathbf{r'})\rho_{ij}(\mathbf{r})}{|\mathbf{r-r'}|}\mathrm{d}\mathbf{r}\mathrm{d}\mathbf{r'}, \label{equ:a1}
\end{equation}
where $\alpha(\mathbf{r, r'})$ is the mixing fraction introduced in Eq.(6) of the main text and $\rho_{ij}(\mathbf{r})=\phi_i^*(\mathbf{r})\phi_j(\mathbf{r})$. The short range exchange energy $E_{x, i}^{\mathrm{SR}}$ is:
\begin{equation}
    E_{x, i}^{\mathrm{SR}} = \sum_j \iint \frac{\mathrm{erfc}\left(\mu(\mathbf{r})|\mathbf{r-r'}|\right)\rho_{ij}^*(\mathbf{r'})\rho_{ij}(\mathbf{r})}{|\mathbf{r-r'}|}\mathrm{d}\mathbf{r}\mathrm{d}\mathbf{r'}. \label{equ: a2}
\end{equation}
Eq.(\ref{equ: a2}) is computationally intractable for an arbitrary function $\mu(\mathbf{r})$. We represent $\mu(\mathbf{r})$ as:
\begin{equation}
    \mu(\mathbf{r}) = \sum_{\mathbf{r_0}}\mu(\mathbf{r_0})\delta(\mathbf{r-r_0}),
\end{equation}
where $\mathbf{r_0}$ is a grid point in real space; Eq.(\ref{equ: a2}) becomes:
\begin{equation}
    E_{x, i}^{\mathrm{SR}} = \sum_{\mathbf{r_0}; j, \mathbf{G}}\frac{\rho^{*}_{ij; \mathbf{r_0}}(\mathbf{G})\rho_{ij}(\mathbf{G})}{|\mathbf{G}|^2}(1-e^{-\frac{|\mathbf{G}|^2}{4\mu(\mathbf{r_0})^2}}), \label{equ: a4}
\end{equation}
where $\rho_{ij}(\mathbf{G})$ and $\rho_{ij, \mathbf{r_0}}(\mathbf{G})$ are the Fourier components of $\rho_{ij}(\mathbf{r})$ and $\rho_{ij}(\mathbf{r})\delta(\mathbf{r-r_0})$, respectively. The computational cost of Eq.(\ref{equ: a4}) is then proportional to the number of grid points ($\mathbf{r_0}$), used to represent $\mu(\mathbf{r})$. \\
\indent We adopt a coarse-grained approximation to the function $\mu(\mathbf{r})$:
\begin{equation}
    \mu(\mathbf{r}) \approx \sum_t \mu_t f_t(\mathbf{r}), \label{equ: 5}
\end{equation}
where:
\begin{equation}
\begin{aligned}
f_t(\mathbf{r}) &= \left\{
    \begin{aligned} 
    & 1 \ \mathrm{if \ } \mathbf{r}\in \Omega_t\\
    & 0 \ \mathrm{else\ where}
    \end{aligned}
    \right.\\
\mu_t &= 1/\Omega_t\int\mu(\mathbf{r})f_t(\mathbf{r})\mathrm{d}\mathbf{r},
\end{aligned}
\end{equation}
 and $\{f_t(\mathbf{r})\}$ are projectors defined in the volume $\{\Omega_t\}$, and $\{\mu_t\}$ are the average of the function $\mu(\mathbf{r})$ in $\{\Omega_t\}$. \\
 \indent The algorithm used here to obtain $\{f_t(\mathbf{r})\}$ and $\{\mu_t\}$, called Adaptive Binning (AB), is the following:

\begin{enumerate}
  \item Evaluate $\mu(\mathbf{r})$ on the real space grid used to represent the charge density.
  \item If the standard deviation ($\sigma$) of $\mu(\mathbf{r})$ is less than $0.1\times\overline{\mu(\mathbf{r})}$ (see below), where $\overline{\mu(\mathbf{r})}$ is the mean value of $\mu(\mathbf{r})$, approximate $\mu(\mathbf{r})$ with $\overline{\mu(\mathbf{r})}$.
  \item If $\sigma$ exceeds $0.1\times\overline{\mu(\mathbf{r})}$:
  \begin{enumerate}
    \item Partition $\mu(\mathbf{r})$ into N=2 bins.
     \item Compute the mean $\{\mu_t\}$ and standard deviation $\{\sigma_t\}$ of $\mu(\mathbf{r})$ in each bin.
     \item Increment the bin number (N) until the following inequality is satisfied:  $|\mu_i - \mu_j| > \sigma_i + \sigma_j$, for any (i, j) bin pairs, where $1 \leq i < j \leq N$.
  \end{enumerate}
  \item For all $\mathbf{r}$, the projector $f_t(\mathbf{r})$ associated to the $t^{\text{th}}$ bin is defined as 1 if $\mu(\mathbf{r})$ belongs to the $t^{\text{th}}$ bin, zero otherwise.The projectors satisfy the following properties: $f_i(\mathbf{r})f_j(\mathbf{r}) = f_i(\mathbf{r}) \delta_{ij}\text{, and } \sum_i f_i(\mathbf{r}) = 1 \forall \mathbf{r} $.\label{step7}
\end{enumerate}

\indent Table (\ref{tab:sigma}) presents the variation of $\mu(\mathbf{r})$ observed in selected homogeneous 3D bulk systems and heterogeneous systems studied in this work. Based on these results, we have chosen $0.1\times\overline{\mu(\mathbf{r})}$ as a reasonable threshold that appears to be sufficiently small to accurately describe variations in heterogeneous systems and not overly sensitive to fluctuations in homogeneous systems.
\begin{table}[htbp!]
    \caption{The ratio between the standard deviation ($\sigma$) and the mean value $\overline{\mu(\mathbf{r})}$ of the function $\mu(\mathbf{r})$ in selected homogeneous and heterogeneous systems.}
\label{tab:sigma}
\renewcommand{\arraystretch}{1.2} 
\tablinesep=5ex\tabcolsep=10pt
\begin{tabular}{lc}  
\toprule
             &   $\sigma/\overline{\mu(\mathbf{r})}$ \\
\midrule
  $\mathrm{Si}$        &  0.06(3)  \\
  $\mathrm{SiO_2}$     &  0.07(1)  \\
  $\mathrm{C}$         &  0.01(9)  \\
  $\mathrm{SiC}$       &  0.03(6)  \\
\midrule
  h-$\mathrm{BN}$        &  0.94(9)  \\
  $\mathrm{Si_{35}H_{36}}$     &  0.94(6)  \\
  $\mathrm{Si/Si_3N_4}$       &  0.08(7)  \\
  H-$\mathrm{Si/H_2O}$      &  0.16(7)  \\
   \bottomrule
\end{tabular}
\renewcommand{\arraystretch}{1} 
\end{table}

\begin{figure}[htbp!]
\includegraphics[width=16cm]{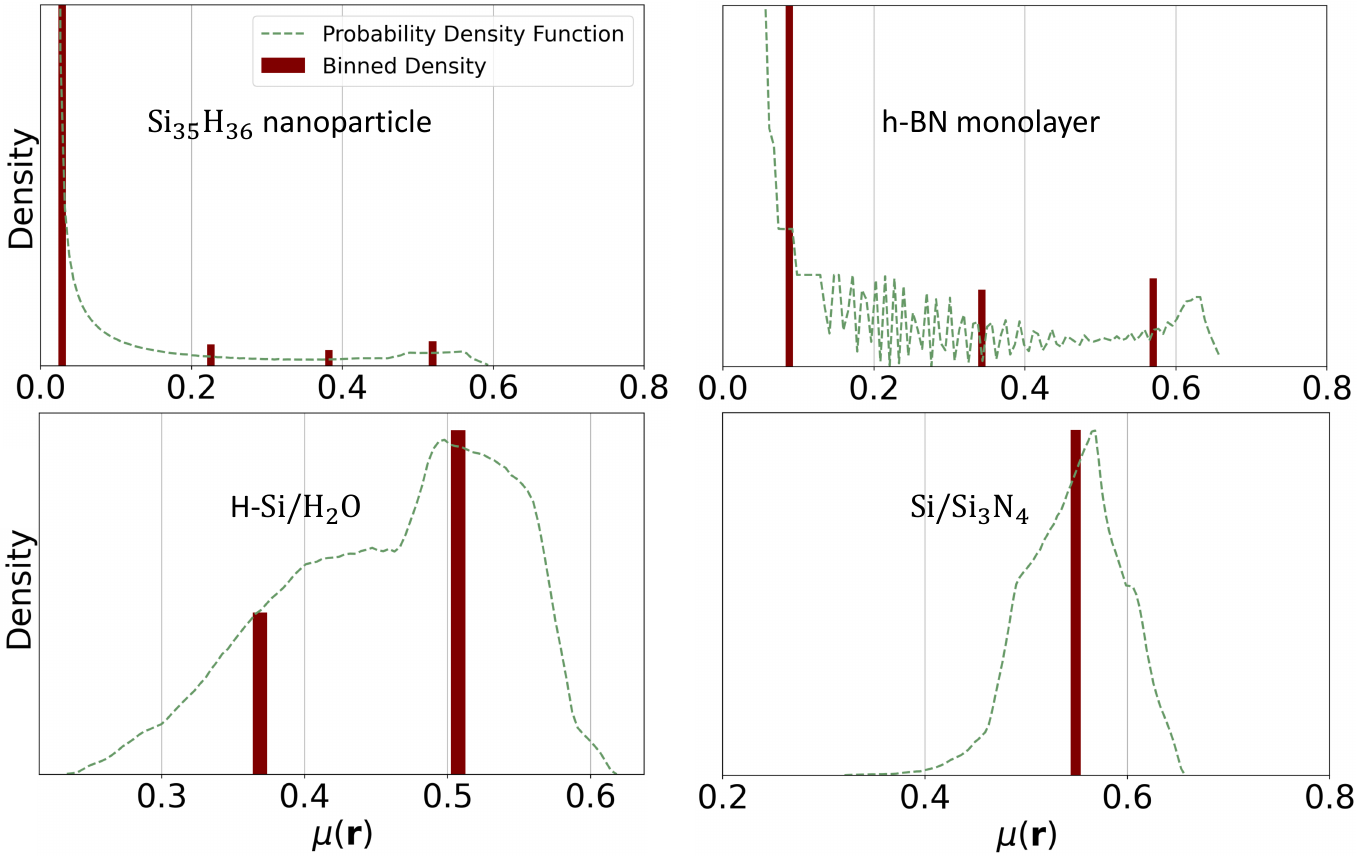}
\caption{The dashed line and histogram show the probability density function and the binned density of $\mu(\mathbf{r})$, respectively, for a $\mathrm{Si_{35}H_{36}}$ silicon cluster, monolayer h-BN, and model H-$\rm{Si/H_2O}$ and Si/$\rm{Si_3N_4}$ interfaces. The bins are determined using the Adaptive Binning (AB) algorithm described in the text.} \label{pic:cg_mu}
\end{figure}
\indent Fig.(\ref{pic:cg_mu}) compares the probability density function and the binned density of $\mu(\mathbf{r})$ for several heterogeneous systems, where bins are determined via the AB algorithm. The choice of how to subdivide the entire supercell into smaller volumes $\Omega_t$ and hence how to determine $\mu_t$ depends on the system. For example, we found that for the Si/$\rm{Si_3N_4}$ interface a single constant parameter $\mu$ is sufficient and hence $\Omega_t$ is equal to the whole supercell. For 2D systems and nanoparticles we found instead that two or more $\mu_t$ paramters are required to obtain an accurate description of the short range component of the exchange and correlation. The use of Eq.(\ref{equ: 5}) was inspired by previous results\cite{10.1103/physrevb.93.235106, 10.1103/physrevmaterials.2.073803} showing that the results of electronic structure calculations with range separated hybrid functionals were relatively insensitive to the choice of $\mu$.\\
 \indent We note that although the complexity of the calculations of the exchange integral of SE-RSH is the same as that of global hybrid functionals', the prefactor is $T$ times larger, where $T$ is the number of $\{\mu_t\}$ used to approximate the function $\mu(\mathbf{r})$. Hence it is desirable to keep $T$ as small as possible, avoiding unnecessary computational complexity.\\
\indent The application of the short-range exchange operator to state $\phi_i$:
\begin{equation}
\begin{aligned}
    (\Sigma_x^{\mathrm{SR}}\phi_i)(\mathbf{r}) = &\sum_j\phi_j(\mathbf{r})\\
    & \times \int\frac{\mathrm{erfc}(\mu(\mathbf{r})|\mathbf{r-r'}|)\rho_{ij}^{*}(\mathbf{r'})}{|\mathbf{r-r'}|}\mathrm{d}\mathbf{r'}
\end{aligned}
\end{equation}
is also evaluated using a coarse-grained approximation of $\mu(\mathbf{r})$.

\section{Divergence in the Evaluation of Exchange Integrals \label{divergence}}
We define the function $F_{ij}(\mathbf{G})$:
\begin{equation}
    F_{ij}(\mathbf{G}) = \iint \alpha(\mathbf{r, r'})\rho_{ij}^*(\mathbf{r'})\rho_{ij}(\mathbf{r})e^{i\mathbf{G\cdot(r-r')}}\mathrm{d}\mathbf{r}\mathrm{d}\mathbf{r'},
\end{equation}
and write the exchange energy $E_{x, i}$ as:
\begin{equation}
E_{x, i} = \sum_{\mathbf{G}, j}\frac{F_{ij}(\mathbf{G})}{|\mathbf{G}|^2}, \label{equ: e_xi}
\end{equation}
The divergence for $\mathbf{G}\rightarrow 0$ is integrated by generalizing the procedure proposed by Gygi et al.\cite{gygi1986self}. An auxiliary function $A_i$ is constructed for each state $\phi_i$ as:
\begin{equation}
    \begin{aligned}
    A_i &= \sum_{G}m_i\frac{e^{-\alpha|G|^{2}}}{|\mathbf{G}|^2}\\
        &= m_i\frac{\Omega}{(2\pi)^{3}} 2\pi\sqrt{\frac{\pi}{\alpha}}
    \end{aligned}
\end{equation}
where $m_i = \sum_j F_{ij}(\mathbf{G}=0)$ and $\Omega$ is the volume of the supercell. Eq.(\ref{equ: e_xi}) can then be rewritten as:
\begin{equation}
\begin{aligned}
        \sum_{\mathbf{G}}\left[\frac{\sum_{j}F_{ij}(\mathbf{G})}{|\mathbf{G}|^2} - m_i\frac{e^{-\alpha|G|^{2}}}{|\mathbf{G}|^2}\right]
         + m_i\frac{\Omega}{(2\pi)^{3}} 2\pi\sqrt{\frac{\pi}{\alpha}},
\end{aligned}
\end{equation}
where 
\begin{equation}
    \left[\frac{\sum_{j}F_{ij}(\mathbf{G})}{|\mathbf{G}|^2} - m_i\frac{e^{-\alpha|G|^{2}}}{|\mathbf{G}|^2}\right]_{\mathbf{G} = 0}
\end{equation}
is a smooth differentiable function. Note that in the original formulation of Gygi et al.'s, $\alpha(\mathbf{r, r'}) = 1$ and hence, $m_i = 1$.
\bibliography{se-rsh}